# Applications of Deep Learning parameterization of Ocean Momentum Forcing

Guosong WANG[1,2], Min HOU[3], Xinrong WU*[1], Xidong WANG*[2], Zhigang GAO[1], Hongli FU[1], Bo DAN[1] , Chunjian SUN[1] , and Xiaoshuang ZHANG[1]

[1]*Key Laboratory of Marine Environmental Information Technology, National Marine Data and Information Service, Tianjin 300171, China*

[2]*The College of Oceanography, Hohai University, Nanjing, Nanjing 211100, China*

[3]*Tianjin Binhai New Area Meteorology Administration, Tianjin 300171, China*

## ABSTRACT

Mesoscale eddies are of utmost importance in understanding ocean dynamics and the transport of heat, salt, and nutrients. Accurate representation of these eddies in ocean models is essential for improving model predictions. However, accurately representing these mesoscale features in numerical models is challenging due to their relatively small size. In this study, we propose a convolutional neural network (CNN) that combines data-driven techniques with physical principles to develop a robust and interpretable parameterization scheme for mesoscale eddies in ocean modeling. We first analyze a high-resolution reanalysis dataset to extract subgrid eddy momentum and use machine learning algorithms to identify patterns and correlations. To ensure physical consistency, we have introduced conservation of momentum constraints in our CNN parameterization scheme through soft and hard constraints. The interpretability analysis illustrate that the pre-trained CNN parameterization shows promising results in accurately solving the resolved mean velocity at the local scale and effectively capturing the representation of unresolved subgrid turbulence processes at the global scale. Furthermore, to validate the CNN parameterization scheme offline, we conduct simulations using the MITgcm ocean model. A series of experiments is conducted to compare the performance of the model with the CNN parameterization scheme and high-resolution simulations. The offline validation using MITgcm simulations demonstrates the effectiveness of the CNN parameterization scheme in improving the representation of mesoscale eddies in the ocean model. Incorporating the CNN parameterization scheme leads to better agreement with high-resolution simulations and a more accurate representation of the kinetic energy spectra.

---
*Corresponding author : Xinrong WU, Xidong WANG
Email: xrw_nmdis@163.com,xidong_wang@hhu.edu.cn

**Article Highlights:**

- We propose a parameterization scheme for mesoscale eddies using a CNN that incorporates interpretability and physical constraints

- The pre-trained CNN parameterization has been successfully integrated into the global ocean model MITgcm for offline testing

- The CNN parameterization is significantly improves the representation of mesoscale eddies and eddy-mean flow interactions in ocean model

**1. Introduction**

The ocean is a complex system that encompasses a wide range of physical processes, from large-scale circulation that spans thousands of kilometers to small-scale turbulence that measures only a few millimeters. Among these processes, mesoscale eddies are of particular importance. These eddies occur at spatial scales of 10-100 km and play a significant role in the surface flow field by accounting for the majority of the kinetic energy (KE) in the ocean. They are crucial for maintaining the global oceanic energy balance and facilitating the energy cascade from large to small scales(Ferrari and Wunsch, 2009). The strong nonlinearity exhibited by mesoscale eddies affects their ability to transport water masses as they propagate(Chelton et al., 2011), thereby influencing climate change through the meridional transport or rotation of heat(Early et al., 2011). However, accurately representing these mesoscale features in numerical models is challenging due to their relatively small size compared to the overall ocean scale. Most ocean models used in the Coupled Model Intercomparison Project (CMIP5 and CMIP6) adopt spatial resolutions between 0.25° and 1°(Eyring et al., 2016), which fail to fully capture physical processes such as oceanic mesoscale eddies(Schneider et al., 2019). This limitation can lead to biased simulations of ocean circulation, ocean heat content, and carbon uptake(Cheng et al., 2017; Farneti et al., 2015).

Ocean models with higher spatial resolution can better simulate mesoscale eddies and air-sea interactions, but they are computationally expensive and also require discretization of spatio-temporally continuous dynamical equations, which leads to unresolved subgrid processes. In numerical models, these unresolved subgrid processes are represented by a simplified function of resolved variables, without considering the details of the physical process. This is called "parameterization". Due to limitations in sampling rates or computational costs(Bauer et al., 2015), global ocean models generally use mesoscale eddy parameterization schemes to approximate subgrid physical processes that cannot be fully resolved, as a way to improve the simulations of large-scale ocean currents and ocean heat uptake(Zhang et al., 2023).

Over the past decades, parameterization schemes based on empirical physical relationships have been the most common and traditional approach in current climate models. These physically driven parameterization schemes generally derive corresponding empirical formulas from relevant physical principles or conservation mechanisms to approximate the overall impact of subgrid processes. Traditional methods for parameterizing ocean mesoscale eddies are based on several key assumptions, such as (1) eddy fluxes of tracers are transmitted downward along a mean gradient, and (2) eddy diffusivity is related to a fixed factor of resolved scale and/or latitude and depth. These assumptions were the basis of the Gent and McWilliams (1990) subgrid eddy parameterization (GM90) (Gent and Mcwilliams, 1990). The GM parameterization aims to parameterize the advective or transport effect of geostrophic eddies by means of a "bolus" velocity. GM90 is well-known for its ability to mimic the effects of baroclinic turbulence, correct spurious diffusion across the isopycnals in the z-coordinate, and improve the simulation results.

The GM-based mesoscale eddy parameterization severely dampens explicit eddies, and the simulations of the global mean mesoscale energy intensity are underestimated by 25%-45% compared to satellite observations(M. Ding et al., 2022). The stochastic parameterization schemes for eddy momentum provide some improvement in simulating the interaction of mesoscale eddies and mean flow, as well as reducing the bias of large-scale ocean currents (Gagne et al., 2020; Guillaumin and Zanna, 2021). A novel KE compensation parameterization method, based on an idealized ocean model, redirects energy back into resolved motions through the application of negative harmonic viscosity on the barotropic velocities(Bachman, 2019). The GEOMETRIC eddy parameterization scheme utilizes a parameterized eddy energy to rescale the GM90 model and shows promise in reproducing the energy balance of mesoscale eddies(Mak et al., 2018; Marshall et al., 2012). Eden and Greatbatch emphasize the importance of isopycnal mixing by mesoscale eddies and propose a new parameterization in which the GM diffusion coefficient is determined by the eddy kinetic energy (EKE), resulting in a more reasonable GM diffusivity diagnosis in the eddy-resolved model(Eden and Greatbatch, 2008). However, it should be noted that due to the significant variation of the first baroclinic Rossby radius of deformation in different regions of the world (below 10 km at high latitudes), even an eddy-resolved ocean model may not achieve consistent eddy simulation performance across all seas. Overall, effectively resolving or parameterizing mesoscale eddies in the gray-zone, where the subgrid process is not well-defined(Gerard et al., 2009), remains challenging and requires further improvement(G. Wang et al., 2021; P. Wang et al., 2021).

Data-driven deep neural networks provide a novel approach to mesoscale eddy parameterization. This approach involves using machine learning or deep learning techniques to directly learn the characteristics of subgrid physical processes from observation or reanalysis data. By automatically searching for optimal parameters using large amounts of training data, this approach eliminates the need for empirical prerequisite physical assumptions. Additionally, it significantly improves computational efficiency, surpassing numerical models by factors of tens or even thousands(Bi et al., 2023; Pathak et al., 2022).The exponential growth of multi-source observations and high-resolution reanalysis products, which contain multiscale spatio-temporal information, provides a valuable resource for enhancing our theoretical understanding of currently unresolved subgrid processes. These data sources can guide the development of data-driven and physics-driven parameterizations(Zhu et al., 2022). Deep learning has demonstrated its potential in improving the parameterization of subgrid processes by accurately extracting complex spatio-temporal features. This has led to advancements in ocean modeling and increased computational efficiency(Li et al., 2020; Zheng et al., 2020) . Data-driven parameterization schemes have shown strong generalization capabilities and can potentially be extended to parameterize other processes(Guillaumin and Zanna, 2021).

However, it is important to note that purely data-driven deep learning models may not always adhere to existing physical principles or constraints, particularly in the absence of sufficient training and validation. Deep learning approaches may also face challenges when applied to domains with different dynamical regimes. Recent research in physics-informed neural networks (PINNs) has emerged to embed relevant knowledge and physical principles into neural network models(Karniadakis et al., 2021). This framework aims to design optimal PINN models that improve the interpretability of neural networks by incorporating physical information constraints during training. By enforcing these constraints, PINNs ensure that the networks comply with the laws of physics and learn more generalizable models with fewer data samples.

Deep learning models are often regarded as black boxes because they have a large number of parameters and layers, making it challenging to explain their behavior(Ghahramani, 2015). Explainable artificial intelligence (XAI) methods have been developed to provide a qualitative understanding of the relationship between extracted features and model outputs (Raissi et al., 2020). However, these XAI methods are primarily focused on RGB images and classification tasks, making them less suitable for high-dimensional scientific data and regression problems (Krishnan, 2020). Despite this, XAI offers a valuable technical foundation to identify shortcomings in data-driven models. Current research on interpretable deep learning primarily examines intrinsic or post-hoc interpretability(Lapuschkin et al., 2019; Samek et al., 2019), model-specific or model-agnostic interpretability, and local or global interpretability(Molnar, 2020). A critical technical challenge for future interpretable deep learning approaches is to meet domain-specific requirements(Rudin, 2019), and advancements in deep learning model interpretability in the field of oceans and meteorology show promising results(Ebert-Uphoff and Hilburn, 2020; McGovern et al., 2019). For instance, Gagne used significant and optimized features to interpret a CNN model for detecting large hail probabilities, uncovering that the CNN neurons preferentially extracted effective environmental and storm morphology information features that are indicative of storms with identifiable morphologies(Gagne II et al., 2019). Toms applied Layer-wise relevance propagation (LRP) to uncover spatial pattern discovery strategies employed by neural networks in climate phenomena analysis(Toms et al., 2020).

The current physics-driven parameterization scheme for mesoscale eddies has limitations, particularly when the grid resolution is not sufficient to resolve these eddies.

In such cases, the large-scale kinetic energy in the model is significantly attenuated. On the other hand, deep learning techniques can efficiently and accurately extract subgrid features that are difficult to solve or overlooked by traditional physics-based methods. However, purely data-driven deep neural networks lack physical constraints and face challenges in terms of interpretation due to their black-box nature.

This study presents a novel approach for capturing mesoscale eddies in ocean modeling by combining data-driven techniques and physical principles. The offline validation using the Massachusetts Institute of Technology general circulation model (MITgcm) (Marshall et al., 1997a, 1997b) ocean model simulations provides strong evidence of the effectiveness of the proposed CNN parameterization scheme. The findings contribute to enhancing the accuracy and reliability of ocean models in representing mesoscale eddies and their role in ocean dynamics and transport processes.

The paper is structured as follows. Section 2 describes the data and analyzes the spatial and temporal distribution of subgrid eddy momentum globally; Section 3 provides a detailed description of the deep learning parameterization approach adopted in this study; Section 4 examines the performance of the global subgrid eddy parameterization in simulations; Section 5 discusses the global interpretability and sample-based interpretability of the deep learning parameterization; Section 6 presents and discusses the results of the MITgcm numerical model with three different mesoscale eddy parameterization schemes. Finally, Section 7 summarizes the findings of the paper.

## 2. Global Subgrid Eddy Momentum: Spatial and Temporal Characteristics

Mesoscale eddies, particularly in regions such as the extensions of the Kuroshio and Gulf Stream in the western boundary currents, as well as in eddy-rich regions like the Antarctic Circumpolar Current in the Southern Ocean, play a pivotal role in heat transport(Abernathey and Marshall, 2013). While numerous studies have concentrated on the time-averaged impacts of eddies on large-scale circulation, it is imperative to take into account the instantaneous variability of eddies, which can be several orders of magnitude larger than the mean(Maximenko et al., 2005) (Maximenko et al., 2005). This section first calculates the global subgrid eddy momentum and analyzes its spatial and temporal characteristics.

### 2.1 Data

The statistical characteristics of mesoscale eddies in the surface layer have been extensively analyzed using satellite altimetry data and drifting buoy observations. However, despite ongoing enhancements in the global ocean observing system, the spatiotemporal coverage of ocean observation data remains relatively limited. Consequently, deriving a quantitative parameterization scheme suitable for scale-aware mesoscale eddies solely from sparse observational data poses a significant challenge. In this study, we employ the Copernicus Marine Environment Monitoring Service global ocean 1/12° physical reanalysis dataset, GLORYS12V1(Jean-Michel et al., 2021), to extract subgrid eddy momentum from the momentum equation using spatial filtering techniques(Bolton and Zanna, 2019; Guillaumin and Zanna, 2021).

The global ocean reanalysis product GLORYS12V1 is derived from the Copernicus Marine Environment Monitoring Service (CMEMS) (https://resources.marine.copernicus. eu/product-detail/GLOBAL_MULTIYEAR_PHY_001_030/). GLORYS12V1 is based on the CMEMS real-time global forecasting system and utilizes the NEMO ocean numerical model driven by ERA-Interim and ERA5 surface data. GLORYS12V1 incorporates assimilation of various observation data sources, including satellite altimetry data, satellite sea surface temperature, sea ice concentration, and in-situ observations of temperature and salinity profiles. The assimilation is performed using the Kalman filter assimilation method. The reanalysis product has a spatial resolution of 1/12°×1/12°, a temporal resolution of daily averages, and covers a time span from 1993 to 2020. It also provides a vertical representation from 0 to 5500 meters below the surface, divided into 50 standard grid layers.

In this study, we construct a subgrid eddy momentum dataset based on the GLORYS12V1 reanalysis product. The dataset is divided into training and test sets. The daily average data from 1993-2019 (a total of 9495 days) is used as the training set, while the data from 2019-2020 (a total of 731 days) is used as the test set for evaluation purposes.

### 2.2 Subgrid eddy momentum

To fix concepts of subgrid eddy momentum, we focus on the momentum equations for a homogeneous incompressible or Boussinesq rotating ideal fluid,

$$\frac{\partial \mathbf{u}}{\partial t} + (\mathbf{u} \bullet \nabla)\mathbf{u} + 2\Omega \times \mathbf{u} + \rho^{-1}\nabla p = \mathbf{F} + D\mathbf{u} \tag{1}$$

$$\nabla \bullet \mathbf{u} = 0 \tag{2}$$

where $\mathbf{u} = (u, v, w)$ is the three-dimensional velocity field, $\Omega$ is the rotation vector, $\rho$ is the constant density, and $p$ is the pressure. The external forcing term $\mathbf{F}$ denotes the curl of the wind stress, which acts only on the surface layer, and $D = \nu\nabla^4\psi - r\nabla^2\psi\delta_m$ is the dissipation term. The fourth-order term $\nu\nabla^4\psi$ denotes the Laplace viscosity, viscosity coefficient of $\nu$ denoting the dissipation of the proposed energy at small scales. The second-order term $r\nabla^2\psi\delta_m$ acts only on the bottom layer, and $r$ is the drag coefficient.

Similar to classical large-eddy simulation, we adopt the Reynolds-averaging method to express the velocity field in the momentum equation. This approach allows us to separate the mean and fluctuating components of the velocity field, which is crucial for understanding and modeling turbulent flows. In this approach, the velocity field in the momentum equation, $\mathbf{u}=\bar{\mathbf{u}}+\mathbf{u}'$, is decomposed into the mean velocity $\bar{\mathbf{u}}$ and the subgrid turbulence $\mathbf{u}'$. The subgrid turbulence $\mathbf{u}'$, which represents the unresolved small-scale features, is incorporated in the subgrid scale stress tensor. The equation suggests that the divergence of the product of filtered thickness and the large-scale mean velocity is counterbalanced by the divergence of the subgrid scale stress tensor. This balance is crucial in maintaining the stability of the system and ensuring accurate simulation results. The subgrid scale stress tensor represents the effects of smaller scales on the larger scales, which are not resolved in the simulation. These effects include turbulent diffusion and transport, which can significantly influence the evolution of the large-scale flow field.

In classical partial differential equations (PDEs), the filtered velocity field $\bar{\mathbf{u}}$ can be obtained by convolving the original velocity field $\mathbf{u}$ with a filter kernel. The filter kernel applies a spatial averaging operation and smoothes out the small-scale variations in the velocity field, resulting in a lower resolution representation. In this paper, the mean velocity field $\bar{\mathbf{u}}$ is obtained by filtering the original velocity field with a Gaussian spatial filter of a filter size (25 km) to capture the large-scale features. This filtering is done to match the resolution of gridded multimission altimeter products. Then $\bar{\mathbf{u}}$ satisfies the equation,

$$\frac{\partial \overline{\mathbf{u}}}{\partial t} + \overline{(\mathbf{u} \bullet \nabla)\mathbf{u}} + 2\Omega \times \overline{\mathbf{u}} + \rho^{-1}\overline{\nabla}\overline{p} = \overline{\mathbf{F}} + \overline{D}\overline{\mathbf{u}} + \mathbf{S}(\mathbf{u}) \tag{3}$$

$$\overline{\nabla}\mathbf{g}\overline{\mathbf{u}} = 0 \tag{4}$$

In this formulation, two simplified assumptions are introduced. Firstly, the influence of Earth's curvature on large-scale motion is considered by accounting for the variation of the Coriolis force term with latitude. The $\beta$- plane approximation is utilized. This approach aids in capturing the effect of Earth's rotation on large-scale oceanic motion. Secondly, given the very small compressibility of seawater, the fluid is assumed to be incompressible using the Boussinesq approximation. This assumption allows for the neglect of variations in density due to minor changes in pressure. However, it should be noted that similar approximate assumptions need to be made for potential temperature and other thermodynamic transport equations when considering the complete equations of seawater motion (Aluie and Kurien, 2011).

By subtracting the equation representing the large-scale mean momentum (Eq.3) from the equation representing the total momentum (Eq.1), we obtain the equation that represents the subgrid eddies momentum $\mathbf{S}(\mathbf{u})$,

$$\mathbf{S}(\mathbf{u}) = (\overline{\mathbf{u}} \bullet \nabla)\overline{\mathbf{u}} - \overline{(\mathbf{u} \bullet \nabla)\mathbf{u}} + \overline{D\mathbf{u}} - \overline{D}\overline{\mathbf{u}} \tag{5}$$

This equation describes the influence of unresolved small-scale features, represented by the fluctuating component of the velocity field, on momentum transport. The subgrid eddies momentum $\mathbf{S}(\mathbf{u})$ represents the effect of these small-scale features on the momentum equation, and therefore needs to be parameterized.

Assuming that the viscosity coefficients in the high-resolution and low-resolution models are different, we introduce the low-resolution coefficient viscosity coefficient as $\nu\overline{\nabla^4\psi}$, the small-scale dissipative term $\nu\overline{\nabla^4\psi}-\nu\overline{\nabla}^4\overline{\psi}$. In this simplified form, the small-scale dissipative term has been neglected since it is one order of magnitude smaller than the term $\mathbf{S(u)}$ (Mana and Zanna, 2014). This approximation assumes that the influence of small-scale dissipation on the subgrid eddies momentum is negligible. Therefore, Eq. (5) can be further simplified to

$$S_U = (S_x, S_y) \approx \overline{(\overline{U} \bullet \nabla)\overline{U}} - \overline{(U \bullet \nabla)U} \qquad (6)$$

### 2.2 Spatial and temporal characterization

The GLORYS12V1 reanalysis product was utilized to examine the spatial and temporal characteristics of global subgrid eddy momentum $S_U$. Figure 1 illustrates the spatial distribution of the global eddy momentum forcing $S_U$, which displays intricate patterns in the surface layer with alternating zonal and meridional directions. The most potent amplitudes are observed around regions such as the Kuroshio and the Gulf Stream and their extensions, as well as the Antarctic Circumpolar Current. Along or across the flow, the $S_U$ exhibits alternating positive and negative potential vorticity characteristics, evolving with the curvature of the flow. The primary effect of the $S_U$ is to redistribute potential vorticity, resulting in alternating gradients of potential vorticity. The eddy-induced velocity accelerates the resolved flow and reduces lateral gradients. The zonal component of the eddy-induced velocity contributes to eddy shedding and participating processes, while the meridional component enhances the western boundary flow, thus increasing the multiscale character of the velocity field(Waterman and Hoskins, 2013).

Figure 2 displays the variability and standard deviation of global zonal surface eddy momentum forcing over the period 2019-2020. The $S_U$ exhibits high spatial and temporal variability, with timescales of several days in eddy-rich regions (Berloff, 2005). The interaction between the mean flow and turbulence of the subgrid eddy momentum improves the structure of the flow, particularly at the western boundary(Li and Storch, 2013). This underscores the importance of considering the spatio-temporal variation of mean and turbulence in the parameterization of subgrid eddy momentum.

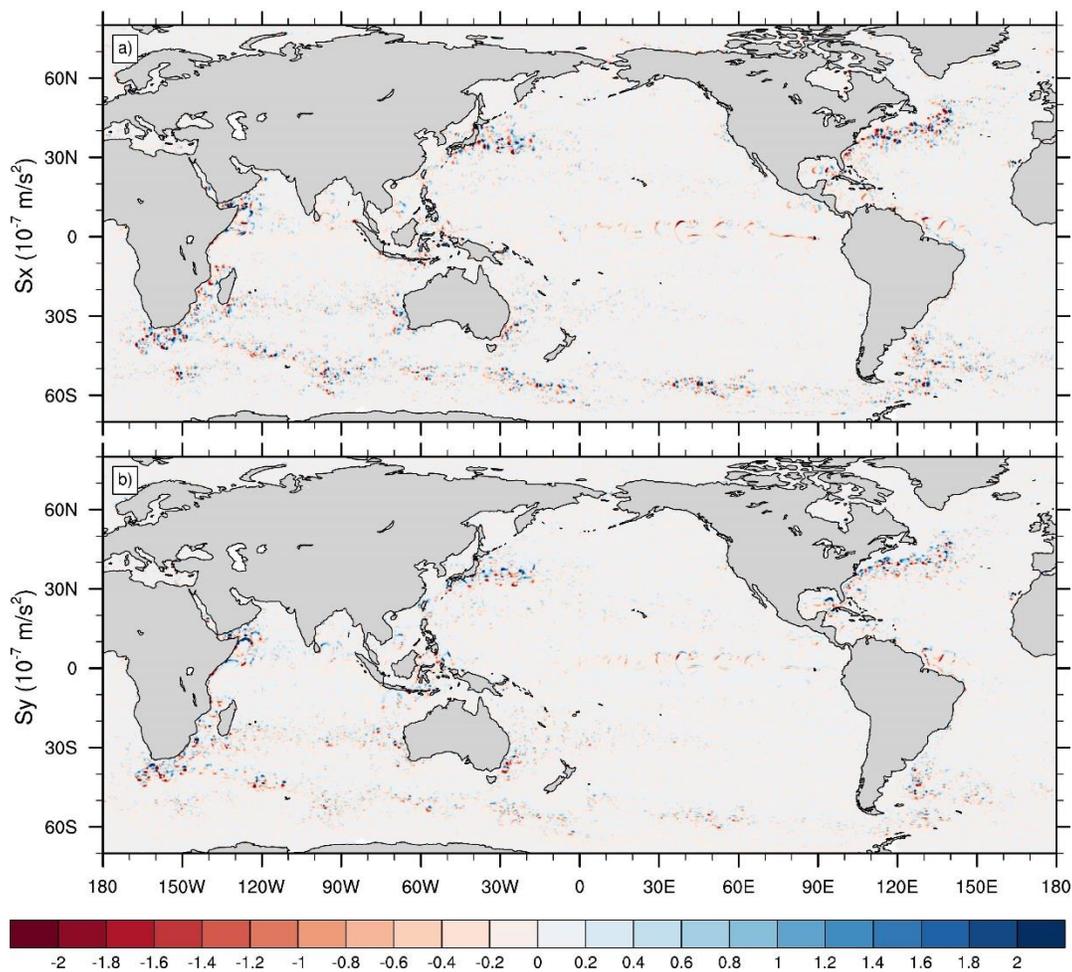

**Fig. 1.** The global zonal (a) and meridional (b) subgrid eddy momentum distributions on January 1, 2019.

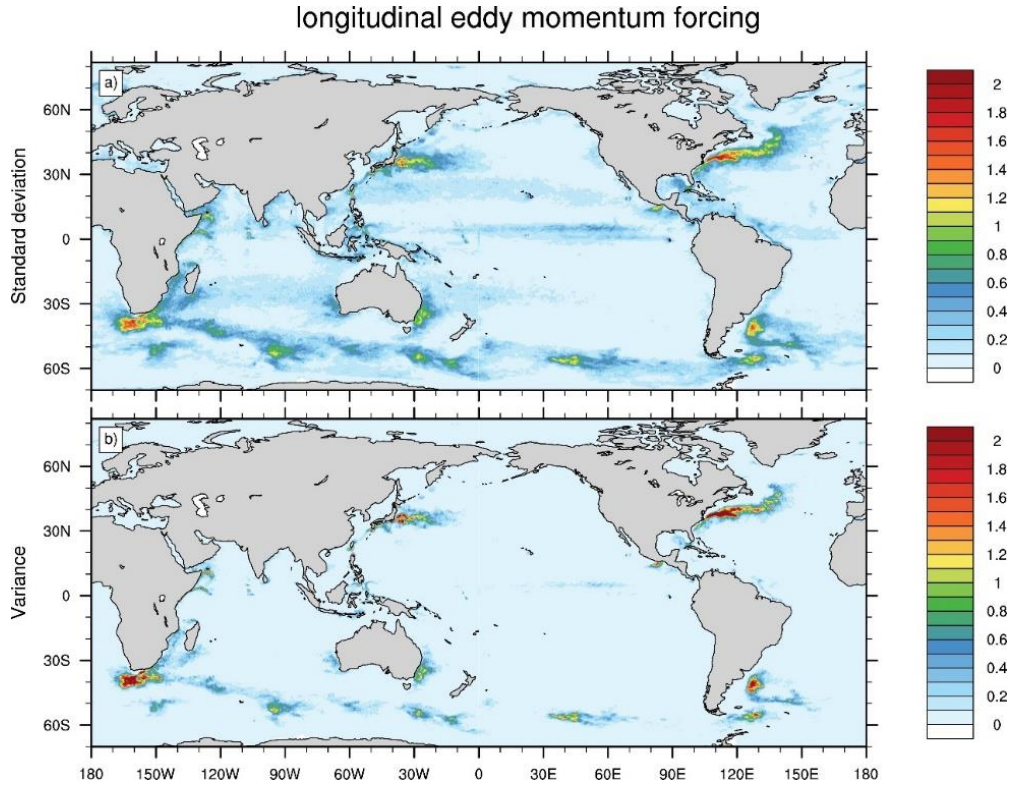

**Fig. 2.** The average global zonal subgrid eddy momentum standard deviation (a) and variance distributions (b) over the surface layer. The data is obtained from the test set and covers the period from 2019-2020.

## 3. Method

### 3.1 *CNN parameterization of subgrid eddies momentum forcing*

In this study, CNN are employed to parameterize the subgrid eddies momentum forcing. CNN have gained popularity in computer vision due to their capability to extract powerful features from input data(Kim et al., 2020). The CNN parameterization architecture utilized in this study, as depicted in Fig.3, comprises five layers with 3×3 convolution kernels to obtain 32, 64, 128, 64, and 32 feature maps. In this architecture, the input grid of size 2041x4320 with 2 variables ($\overline{u}$ and $\overline{v}$) is passed through five convolutional layers to extract relevant features. These convolutional layers apply filters to the input grid and produce feature maps. The feature maps are then flattened and passed through fully connected layers (FC) which perform further computations and transformations. Finally, the output layer generates a grid of size 2041x4320 with 1 variable ($S_U$). The blue dashed box were later used for represents the hard-constrained component of the network, which imposes certain constraints and restrictions on the model's learning process. These constraints help in maintaining the accuracy and reliability of the network's predictions for global subgrid eddy momentum.

To accommodate the distinct spatial and temporal characteristics of the global zonal and meridional subgrid eddy momentum forcing, two separate CNN parameterization are constructed. One is designated for the zonal subgrid eddy momentum forcing, denoted as $f_x = (\overline{U}, \omega_x)$, while the other is for the meridional subgrid eddy momentum forcing, denoted as $f_y = (\overline{U}, \omega_y)$. The inputs to these CNNs are the resolved surface velocity field, denoted as $\overline{U}$, and the outputs are the corresponding subgrid eddy momentum forcing, denoted as $S_U = (S_x, S_y)$. The CNN parameterization facilitates the extraction of relevant features from the resolved surface velocity field to estimate the subgrid eddy momentum forcing. By training separate CNN parameterization for the zonal and meridional components, the different spatial distribution characteristics of the subgrid eddy momentum forcing can be adequately captured.

Given that the eddies are concentrated at the surface, and the amplitude of these small-scale eddies decreases rapidly with increasing depth, the CNN parameterization focuses solely on the surface layer. The CNN parameterization are trained using the Keras library (Chollet, 2015) with a TensorFlow backend(Abadi et al., 2016). The training process involves a minimum of 300 epochs on NVIDIA Quadro P6000 GPUs with 50G Memory. Each CNN parameterization has a total of 185,505 parameters. (See Appendix A).

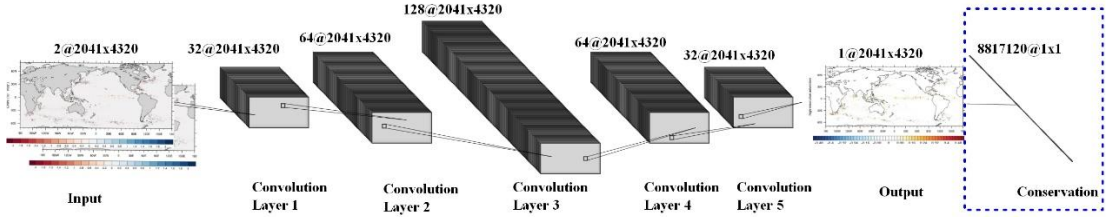

**Fig. 3.** The schematic representation of the architecture of the Hard-constrained (blue dashed box) CNN parameterization for global subgrid eddy momentum. Each input example consists of a 2041x4320 grid of 2 variables, and the output is a 2041x4320 grid of 1 variable.

### 3.2 *Global momentum conservation constraint*

The conservation law of global subgrid eddy momentum posits that the spatial average of subgrid eddy momentum in subregions can have a non-zero tendency, but the global spatial average should be zero. Therefore, it is crucial to ensure that the CNN parameterization subgrid eddy momentum forcing, $\hat{S}_U = (\hat{S}_x, \hat{S}_y)$, by the deep neural networks ($f_x$ and $f_y$) do not introduce spurious momentum sources or sinks, and that the global net momentum adheres to the physical conservation law ($\nabla g \hat{S}_U = 0$).

To achieve this, two methods, namely "soft constraint" and "hard constraint" are employed to incorporate the global momentum conservation into the neural networks, thereby ensuring the resulting parameterization is not only data-driven but also physically meaningful and consistent with the overall dynamics of the ocean system.

The "soft constraint" method is commonly used to enforce physical constraints by enhancing the loss function through the addition of relevant constraints. In this study, the soft constraint involves adding the predicted global eddy momentum conservation loss function, $L_c = N^{-1} \left| \sum_{i=1}^{N} \hat{S}_i \right|$, to the original loss function $L = N^{-1} \sum_{i=1}^{N} (S_i - \hat{S}_i)^2$ (Eq. (7)). This means that the difference between the predicted and the ground truth values of global momentum before and after each iteration is taken into account in the loss function of the neural network. By doing so, the conservation law "participates" in the training process and helps prevent the introduction of spurious momentum sources or sinks.

Implementing the soft constraint approach ensures that the neural networks actively maintain the physical conservation of global subgrid eddy momentum during the training process. This helps improve the accuracy and reliability of the predicted subgrid eddy momentum forcing while adhering to the physical principles of conservation.

$$Loss = L + L_c = N^{-1} \sum_{i=1}^{N} (S_i - \hat{S}_i)^2 + N^{-1} \left| \sum_{i=1}^{N} \hat{S}_i \right| \tag{7}$$

While the soft constraint method is effective in extracting features, it may not strictly satisfy the constraint conditions in different dynamical regimes. In contrast, the "hard constraint" approach involves customizing the neural network structure based on the physical characteristics, allowing for strict adherence to physical constraints in different dynamical regimes. This approach is robust and is considered a favorable method for enforcing physical constraints(Chen and Zhang, 2020; Xu and Darve, 2022).

To implement the hard constraint, the design of the CNN parameterization is modified to ensure that the desired physical constraints are met. This can involve adding specific layers or connections to the network architecture to enforce the constraints in different dynamical regimes. In this study, in the context of subgrid eddies momentum forcing, a physically conservative structure is constructed by introducing a FC to the top layer of the CNN parameterization. This modification is depicted in Fig. 3 and outlined in Eq. (8). This FC can be designed to subtract the output of each grid point from the spatial average. This adjustment helps reduce net momentum bias and ensures strict adherence to the conservation law.

$$\hat{S}_i = \hat{S}_i - \frac{1}{N} \sum_{i=1}^{N} \hat{S}_i \tag{8}$$

The advantage of the hard constraint method is that it explicitly incorporates the physical constraints into the network architecture, ensuring the desired properties are maintained throughout the training and prediction process. This approach can provide more robust and accurate results, particularly in scenarios where the soft constraint method may not be sufficient to strictly adhere to the physical constraints.

## 4. Evaluation

### 4.1 *Assessment of physical conservation constraints*

The assessment of physical conservation constraints is conducted in this section. Figure 4 presents the analysis of global spatially average zonal and meridional eddy momentum deviations. Fig. 4a demonstrates the time series of zonal subgrid eddy momentum in the middle of the Kuroshio Extension. It reveals that the subgrid eddy momentum exhibits prominent daily variation characteristics in this region. The maximum amplitude variation interval spans from -3.7 to 3.8 m/s$^2$. Notably, the CNN parameterization results $\hat{S}_U$ obtained from the CNN parameterization closely align with the high-resolution reanalysis $S_U$. Fig. 4b illustrates the time series of global average zonal subgrid eddy momentum. Due to truncation errors, the high-resolution reanalysis exhibit a small deviation from zero, with a relative error of: $o(10^{-3})$. In contrast, the CNN parameterization strictly align with zero. This achievement can be attributed to the incorporation of the "soft constraint" and "hard constraint" methods, which ensure the adherence to global physical conservation in the neural network.

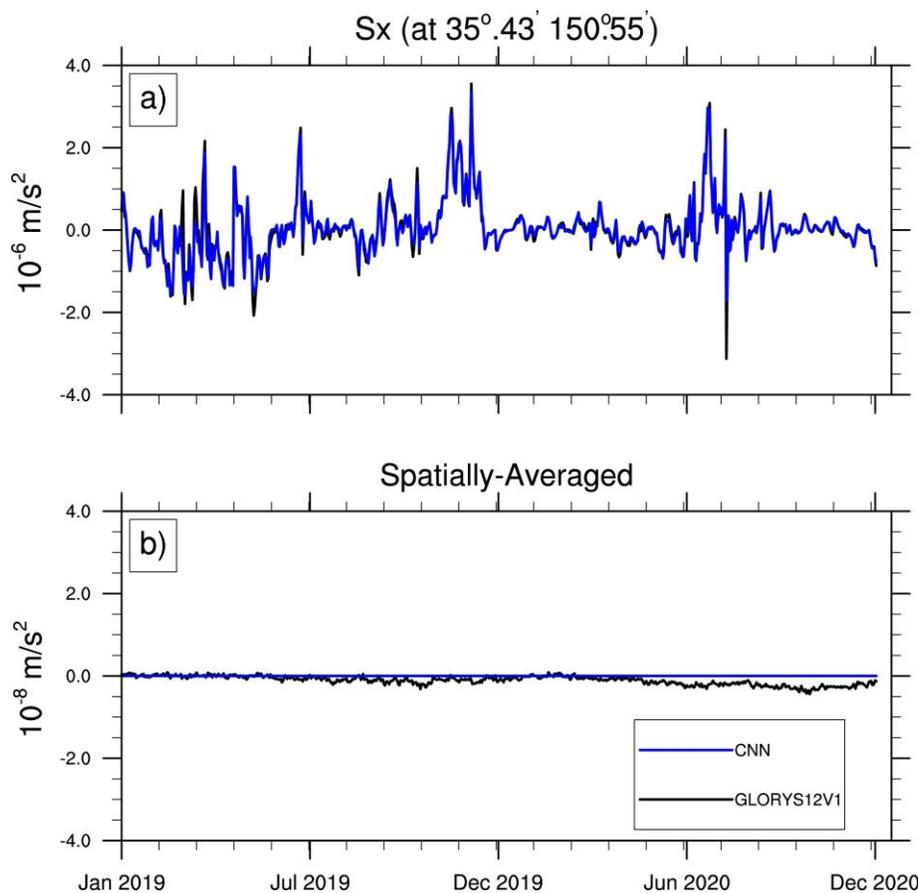

**Fig. 4.** Panel (a) presents the time series of the zonal components of the subgrid momentum forcing at 35°43´, 150°55´, located near the middle of the Kuroshio Extension. Panel (b) displays the time series of the zonal components of the global average momentum forcing across the entire globe. The data is obtained from the test set and covers the period from 2019-2020.

### 4.2 *Global subgrid eddies momentum forcing*

The spatiotemporal variability, Pearson correlation coefficient (PCC) and root-mean-square error (RMSE) of $S_U$ and $\hat{S}_U$ from the high-resolution reanalysis and CNN parameterization are shown in Fig.5 and Fig.6. Figure 5 illustrates the agreement between the global spatiotemporal variability patterns of the CNN parameterization $\hat{S}_U$ and the high-resolution reanalysis $S_U$ . It can be observed that the CNN parameterization successfully reproduces the amplitude and variability of the subgrid eddy momentum, demonstrating excellent agreement with the high-resolution reanalysis. Figure 6 further evaluates the accuracy of the CNN parameterization by examining the RMSE and PCC between the CNN parameterization and the high-resolution reanalysis of the meridional and zonal subgrid eddy momentum. The RMSE values for both components are nearly zero, indicating that the CNN parameterization match the high-resolution reanalysis with high precision. The PCC values are above 90%, indicating a strong correlation between the CNN parameterization and the high-resolution reanalysis. Furthermore, it is worth noting that the meridional deviation is smaller than the zonal deviation, suggesting that the CNN parameterization effectively captures the spatial characteristics of the subgrid eddy momentum.

In summary, the CNN parameterization successfully simulates the spatiotemporal variability of unresolved global subgrid eddy momentum in various regions, demonstrating well agreement with the high-resolution reanalysis. Additionally, the model adheres to the physical constraint of momentum conservation. These findings highlight the capability of the CNN parameterization to accurately parameterize and reproduce the behavior of subgrid eddies using only low-resolution velocity data as inputs.

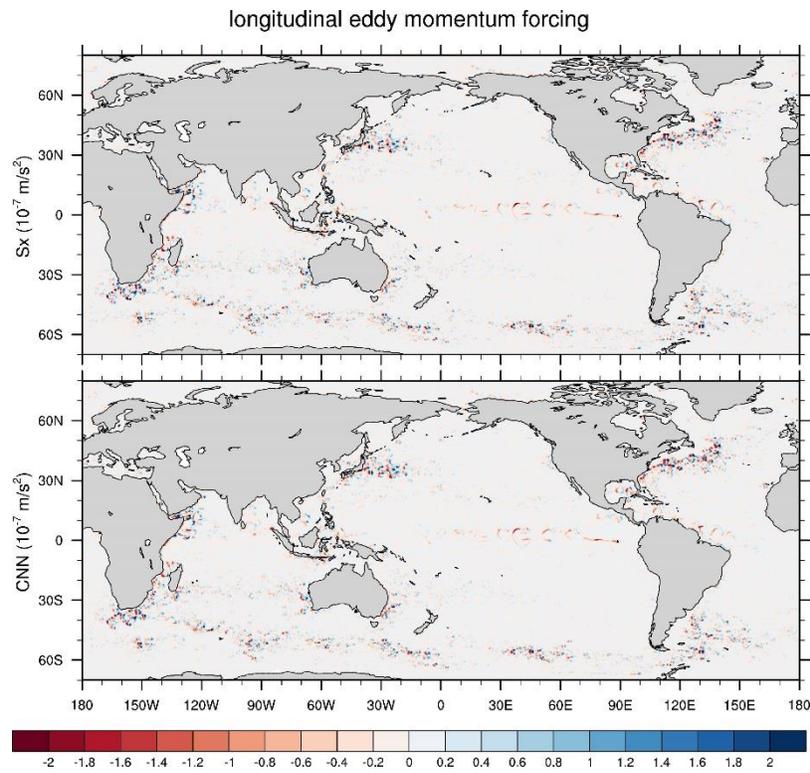

**Fig. 5.** High-resolution reanalysis (a) and CNN parameterization (b) of the global zonal subgrid eddy momentum on January 1, 2019.

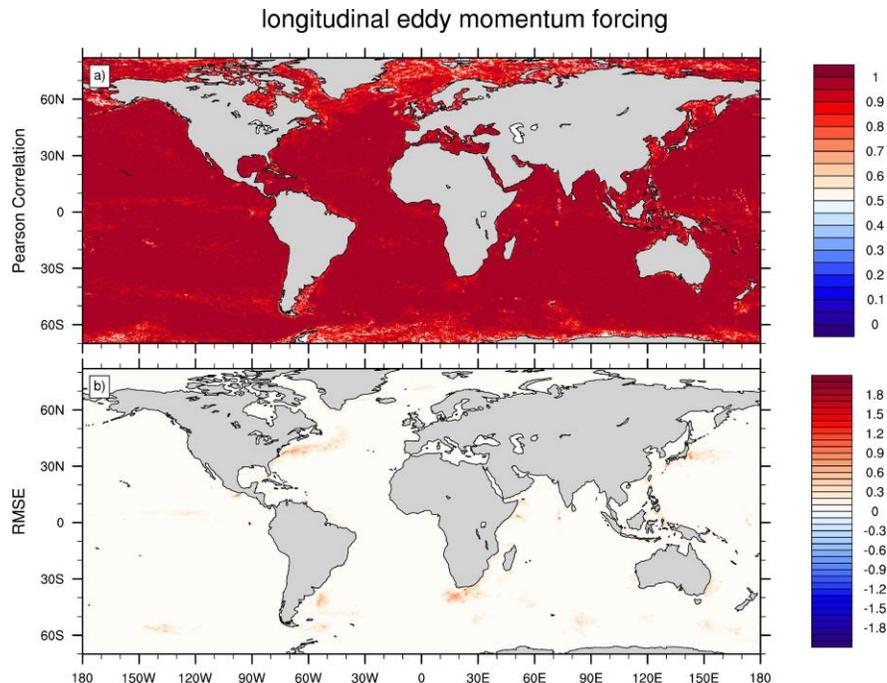

**Fig. 6.** The distribution of the average PCC (a) and RMSE (b) for the global meridional subgrid eddy momentum based on CNN parameterization. The data is obtained from the test set and covers the period from 2019-2020. Higher PCC values suggest a stronger correlation between the CNN parameterization and the high-resolution reanalysis data, indicating a better reproduction of the meridional subgrid eddy momentum patterns. Lower RMSE values indicate a smaller average difference between the simulations and the high-resolution reanalysis data, suggesting a higher accuracy and reliability of the CNN parameterization in capturing the meridional subgrid eddy momentum variations.

## 5. Interpretability analysis

Neural networks, particularly deep neural networks, have the ability to effectively capture complex patterns in vast amounts of data due to their large number of parameters. However, achieving interpretability in these models can be challenging. Methods such as visualization techniques and feature attribution methods are being developed to gain insights into the interpretations made by these models. These approaches aim to uncover the meaningful features learned by the CNN, facilitating the understanding and interpretation of their decision-making processes.

## 5.1 *Global interpretation*

CNN commonly used in image analysis tasks, employ features that represent various properties of the data, enabling description and understanding of the information. These features can be broadly categorized into low-level and high-level attributes(Li and Yu, 2016). Low-level attributes are typically extracted at a local scale and are sensitive to small changes in the input. They capture details and fine-grained information within specific regions of the input. On the other hand, high-level attributes are derived at a global scale, taking into account the entire input domain. These features are more robust and provide a broader understanding of the data.

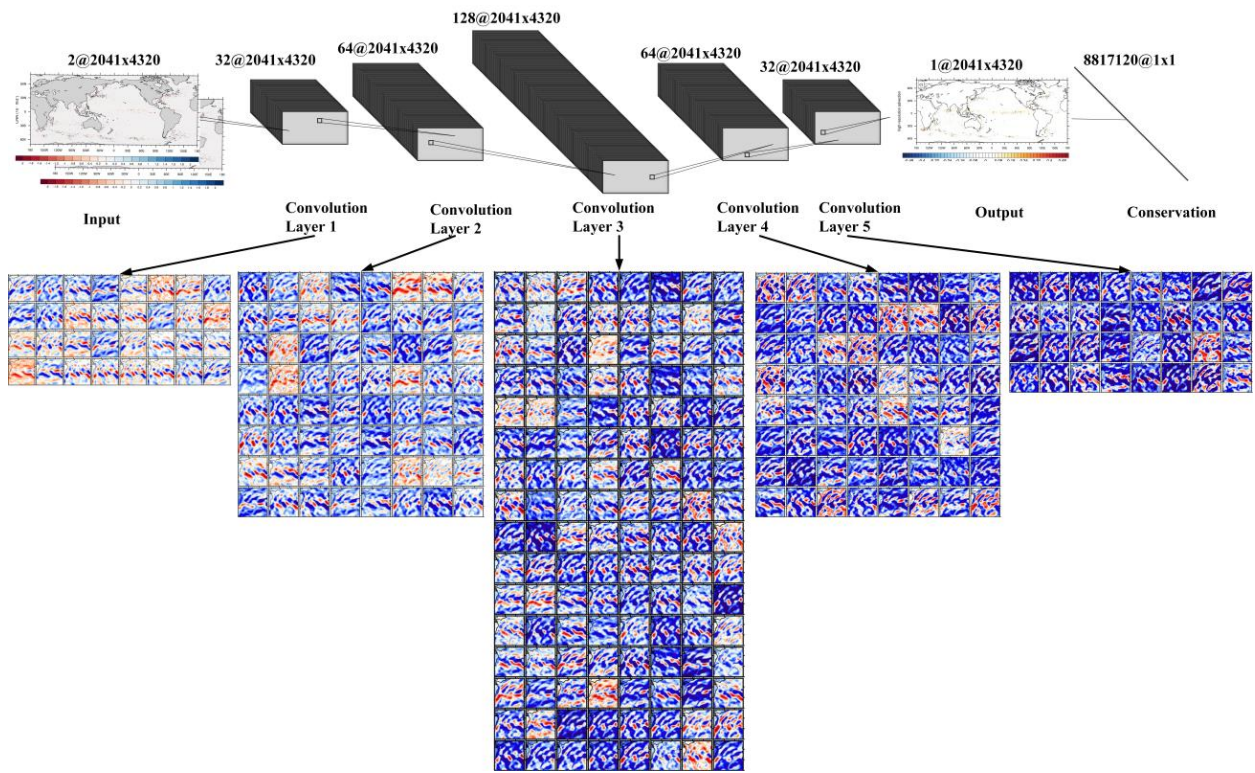

**Fig. 7.** The architecture of the CNN parameterization for global subgrid eddy momentum including convolutional layers and feature maps. The CNN parameterization takes two grid variables related to the subgrid eddy momentum as input, and the first convolutional layer convolves the input variables with a set of filters. Each filter extracts specific features or patterns from the input data. Feature maps are the output of each convolutional layer. They represent the intensity of activation of specific filters across the input. To visualize these feature maps, a color scheme is commonly used, with negative values shown in blue and positive values displayed in red. Each subsequent convolutional layer further refines the extracted features, resulting in a multi-layer representation of the input.

Figure 7 illustrates the feature maps obtained from the CNN parameterization. These feature maps contain rich information, but their interpretation can be complex. It is important to note that there may be correlations among the features, and analyzing each feature individually may not effectively utilize the information present in the feature maps(Montavon et al., 2018).To address this, this paper employs the empirical orthogonal functions (EOF) to decompose the feature maps obtained from the five convolutional layers. The use of EOF allows for a more meaningful interpretation of the features by identifying the dominant spatial patterns present in the data. Additionally, the paper utilizes a set of spatial derivative functions of the velocity field, as proposed by Pope (1975) and Smagorinsky (1963) in Equation 9 (Pope, 1975; Smagorinsky, 1963),

$$SPD = \sqrt{u^2 + v^2}$$
$$RV = v_x - u_y$$
$$DV = u_x + v_y$$
$$STD = v_x + u_y \qquad (9)$$
$$SHD = u_x - v_y$$
$$TD = \sqrt{STD^2 + SHD^2}$$

where SPD, RV, DV, STD, SHD and TD denote speed, relative velocity, divergence, stretch-deformation, shear-deformation, and total-deformation, respectively. It is worth noting that incorporating these decomposition techniques and spatial derivatives assists in deciphering and understanding the representation of subgrid eddies momentum within the CNN parameterization. By analyzing the dominant spatial patterns and employing derivative functions, a deeper insight into the parametric representation of the subgrid eddies momentum can be gained. These functions help reveal the parametric representation of the subgrid eddies momentum.

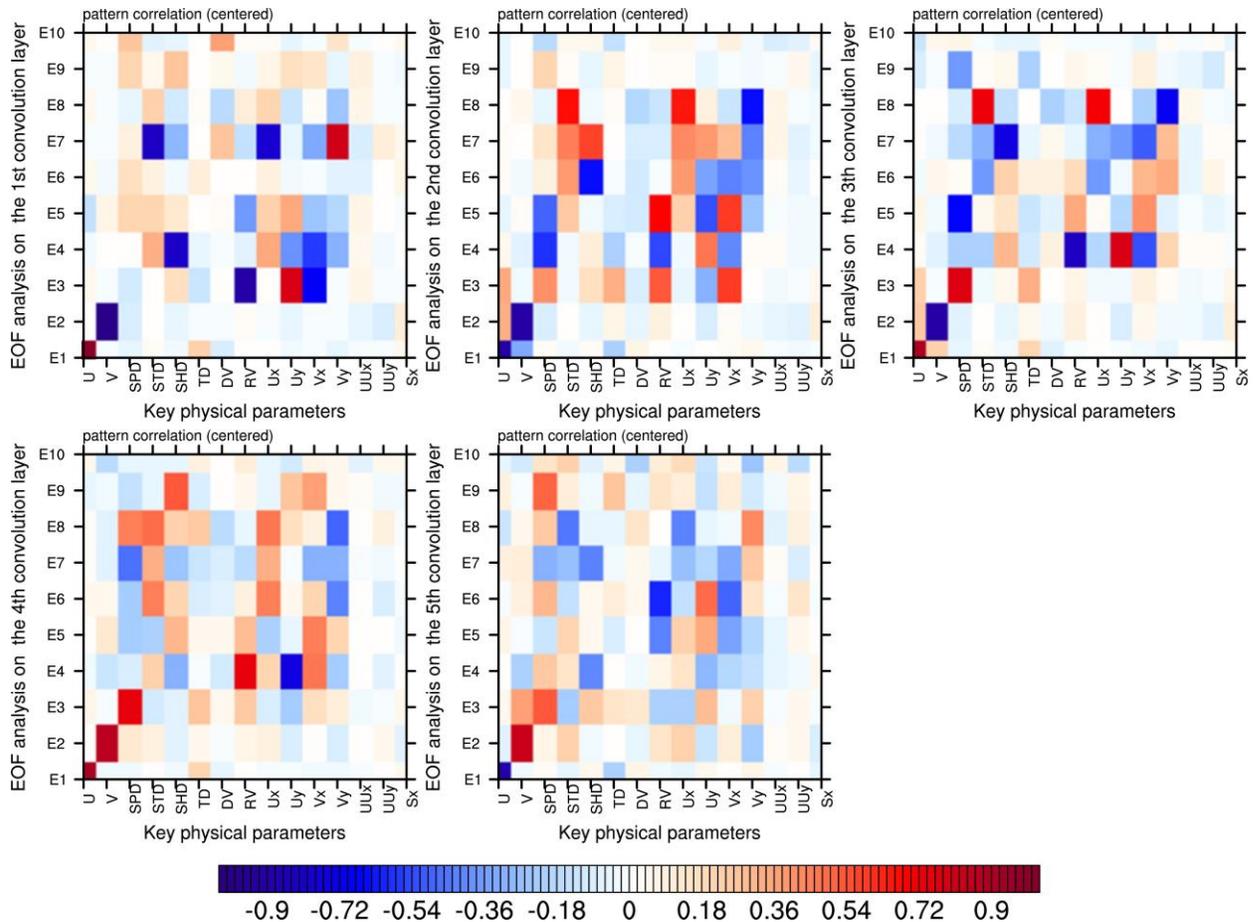

**Fig. 8.** The relationship between the key physical parameters and the first ten spatial modes of the EOF decomposition in the Kuroshio Extension dynamics on January 1, 2019. The first ten spatial modes extracted from the EOF decomposition represent the most significant coherent structures or patterns of variability observed in the Kuroshio Extension on that date.

The correlation analysis between the main extracted features from the CNN parameterization and key physical processes in Equation 8 is presented in Fig. 8. The results demonstrate that the low-level features extracted from the shallow convolutional layers primarily focus on the input zonal and meridional velocity fields (U, V) and their corresponding gradients. These features exhibit a strong correlation with the key physical processes, with correlation coefficients exceeding 80%. As the convolutional layer depth increases, a larger number of features are extracted from the deep convolutional layers. This suggests that the CNN parameterization is able to capture more high-level features (STD, SHD, and TS), which are crucial for simulating subgrid eddy momentum. By combining low and high-level information, the CNN parameterization can achieve high accuracy and comprehension in processing inputs. The integration of both types of features allows the network to capture and analyze fine details as well as global patterns, resulting in a comprehensive understanding of the subgrid eddy momentum.

Overall, the global interpretability analysis reveals that the capability of the CNN parameterization to extract meaningful features that are closely linked to the key physical processes involved in the simulation of subgrid eddy momentum. The strong correlations observed between the extracted features and physical processes validate the effectiveness of the CNN parameterization in capturing and representing these important dynamics. However, it is important to note that our analysis has focused on known features, and there may be other features extracted by the CNN parameterization that are related to unknown key physical processes. Further analysis is required to investigate these unknown features and their implications.

### 5.2 *Example-based interpretation*

Example-based explanations, also known as instance-based explanations, are widely used in machine learning to interpret the results of deep learning models and understand the complex underlying data distributions. This approach involves selecting specific instances from the dataset to provide explanations for how the model operates. There are several popular example-based explanation methods, including saliency maps (Simonyan et al., 2013), SmoothGrad (Smilkov et al., 2017), PatternNet and PatternAttribution (Kindermans et al., 2018), LRP (Binder et al., 2016), and DeepTaylor (Montavon et al., 2017), among others. Each of these methods makes different assumptions and addresses specific objectives(Montavon et al., 2018).

LRP is a powerful neural network visualization method that can provide a comprehensive understanding of the information flow within the network. LRP is used to uncover spatial patterns related to climate in neural networks (Barnes et al., 2019) or identify convective phenomena in satellite images (Lee et al., 2018). In LRP, specific backpropagation rules, such as the ε-rule or β-rule, are tailored to propagate the regression results backward through the neural network. By tracing back from the output neuron to the input features, LRP calculates the relevance estimation between the output neuron and the input features. This integration of interpretability into complex deep learning neural networks is a significant aspect of LRP. Since LRP is a complex research topic, the details of its implementation and algorithms are beyond the scope of this paper. However, for a comprehensive understanding, readers are encouraged to refer to Binder et al. (2016) for a detailed introduction to LRP.

In this study, our main objective is to gain a deeper understanding of the CNN parameterization. To do so, we apply interpretable analyzers specifically designed for regression tasks to a specific region of interest, in this case, the Kuroshio Extension (147°E-152°E, 36°N-41°N). The example-based explanation methods are originally designed to work with regression tasks, where the output neurons represent different parameterized simulation results. However, since our goal is to understand the underlying physical logic of the model, we convert the two-dimensional tensor output into a one-dimensional vector. By identifying the maximum value within this vector, we can determine the interpretable target.

To analyze the subgrid eddy momentum parameterization, we utilize a pre-trained CNN parameterization and input samples from the region of interest. Using the pixel coordinates of the maximum output value as a reference point, we apply three different visualization interpretable analyzers (SmoothGrad, DeepTaylor and LRP). Specifically, we generate heat maps that visualize the maximum values of the parameterized subgrid eddy momentum across all output neurons for five different spatiotemporal features. This analysis provides a comprehensive evaluation of the three visualization interpretable analyzers and their ability to accurately interpret and explain the resolved subgrid nonlinear advection terms in the given examples.

Meanwhile, the global subgrid eddy momentum, as calculated in Eq.(6), can be expanded and written in detail as follows,

$$S_U = \begin{pmatrix} Sx \\ Sy \end{pmatrix} = (\overline{U} \, g \overline{\nabla}) \overline{U} - \overline{(U \, g \overline{\nabla}) U} \tag{10}$$

where

$$Sx = (\overline{\overline{u}\overline{u}}_x + \overline{\overline{v}\overline{u}}_y) - \overline{(uu_x + vu_y)}$$
$$Sy = (\overline{\overline{u}\overline{v}}_x + \overline{\overline{v}\overline{v}}_y) - \overline{(uv_x + vv_y)} \tag{11}$$

Eq. (11) can be decomposed into two terms: the resolved nonlinear advection term (first term on the right-hand side) and the average of the unresolved subgrid nonlinear advection terms (second term on the right). Figure 9 illustrates the results of the five example-based interpretable analyses for the resolved nonlinear advection terms. The label indexes displayed on the left and right sides of each row indicate that the CNN parameterization closely matches the high-resolution reanalysis, indicating the reliability of the interpretable analysis.

The interpretation of the pre-trained CNN parameterization algorithms is represented through heat maps. The heat maps generated by all three visualization interpretable analyzers in Fig.9 are concentrated around the local pixel with the maximum output value. This suggests that the easiest way to increase the output pixel value is by increasing the gradients. The pre-trained CNN parameterization successfully learns the strong nonlinearity between the subgrid eddy momentum and the velocity fields, as well as their gradients. Furthermore, the SmoothGrad method, which is based on the standard gradient, produces heat maps that are more concentrated in the local domain of the maximum output pixel in all five examples. This indicates that the pre-trained CNN parameterization effectively extracts the local nonlinear features of the resolved nonlinear advection terms through the convolution kernels.

It is important to note that the features are extracted by the convolution kernels of each subdomain, typically with a size of 3×3. The LRP heat maps demonstrate that the positive and negative features of the CNN parameterization are similar to the numerically solved resolved nonlinear advection terms (as shown in the fourth column of Fig.9). These different visualization interpretable analyzers collectively demonstrate that the pre-trained CNN parameterization has the ability to accurately solve the resolved mean quantities.

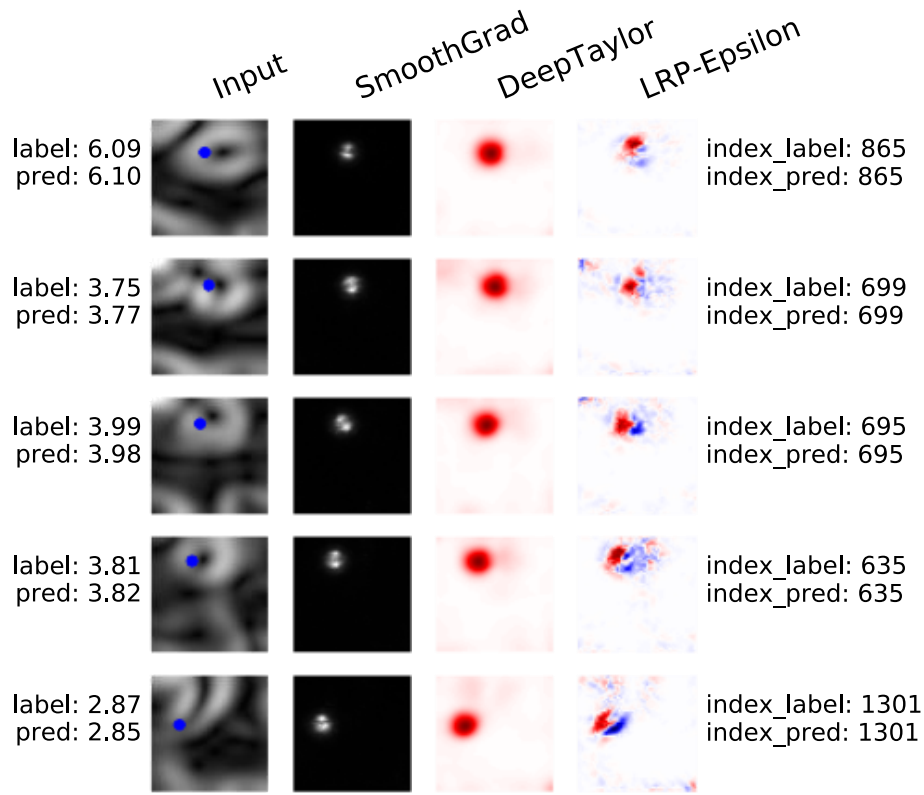

**Fig. 9.** An analysis was conducted to compare and interpret three different visualization interpretable analyzers for resolving subgrid nonlinear advection terms. The analysis involved five typical examples (January 1, 2019, July 23, 2019, January 7, 2020, May 12, 2020, October 9, 2020), with each column displaying the visualized results produced by different visualization interpretable analyzers. Each row represents the analysis for a specific input example. On the left side of each row, the ground truth value and the CNN parameterization value are presented. On the right side, the ground truth label index and the CNN parameterization label index are provided. These indices indicate the corresponding labels assigned to the input examples. The interpretation of the CNN parameterization algorithms is represented through heat maps. The iNNvestigate package (Alber et al., 2019) was employed to calculate the LRP and other heat maps. In the case of LRP, the $\alpha$-$\beta$ rule was applied with $\alpha = 1$ and $\beta = 0$.

In contrast, the results of the unresolved nonlinear advection term in Figure 10 show that all three visualization interpretable analyzers not only highlight local correlations but also focus on a global scale in the heat maps when compared to the resolved nonlinear advection terms in the five examples. This suggests that the range of unresolved subgrid features not only extracts local gradient features around the maximum output pixel but also successfully captures distant nonlinear features at a global scale in regions of unresolved strong advection velocity. This is achieved through successive convolutional layers with a large effective receptive field(Z. Ding et al., 2022).

Moreover, the pre-trained CNN parameterization, using resolved velocity field inputs, reflects the features of the unresolved subgrid, indicating that the pre-trained CNN parameterization possesses good simulation generalization abilities for unresolved subgrid turbulence processes. It is worth noting that the heat maps in the five examples also exhibit differences, indicating that the CNN parameterization utilizes different feature parameters for examples with different spatio-temporal features. This observation forms the basis for the scale-aware CNN subgrid eddy momentum parametric model.

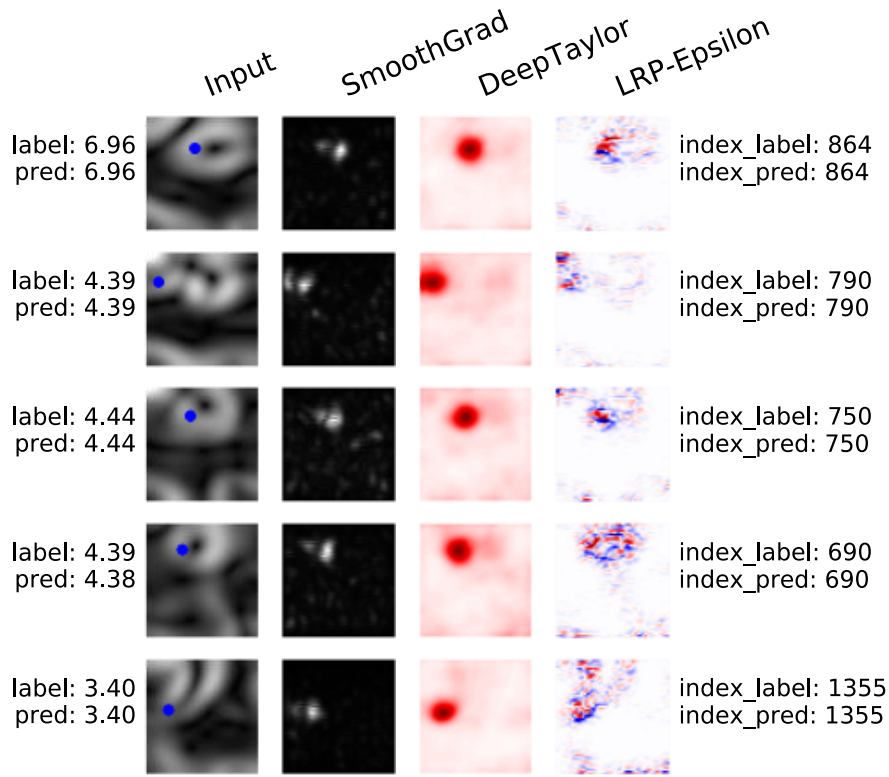

**Fig. 10.** Same as Fig.9, but for the unresolved subgrid nonlinear advection terms.

In the mathematical theory of artificial neural networks, universal approximation theorems state that feedforward neural networks with a single hidden layer and a finite number of neural units can approximate continuous functions(Cybenko, 1989; Hornik et al., 1989). Additionally, deep learning techniques, which utilize reverse mode automatic differentiation for gradient solving, have demonstrated that deeper networks are more likely to learn optimal solutions and reduce generalization errors by virtue of the powerful fitting capabilities (Goodfellow et al., 2016). Overall, the example-based interpretation results illustrate that the pre-trained CNN parameterization shows promising results in accurately solving the resolved mean velocity at the local scale and effectively capturing the representation of unresolved subgrid turbulence processes at the global scale. This suggests that the pre-trained CNN can accurately simulate and represent both mean velocity and subgrid turbulence aspects of the problem.

## 6. Implementation into the MITgcm ocean model

One of the main challenges in incorporating deep learning parameterizations into ocean modeling is ensuring their feasibility and stability under different conditions (Brenowitz and Bretherton, 2018). The integration of deep learning with ocean numerical models still requires further development (Gentine et al., 2018). In this section, we integrate the pre-trained CNN parameterization scheme into the MITgcm numerical model (MITgcm+CNN framework) to evaluate the kinetic energy backscatter and offline generalization properties of the parameterization.

The eddy-resolved MITgcm numerical model used in this study employs a horizontal grid based on a cube-sphere grid projection. The grid is divided into six faces, with each face consisting of 1020×1020 grid cells and a horizontal resolution of approximately 9 km. For the vertical levels, a 50-layer Z-coordinate system is used. The atmospheric forcing for the model includes 10m wind velocity, air temperature, specific humidity, precipitation, downward longwave radiation, and shortwave radiation. These variables are obtained from the 55-year Japanese Reanalysis Project (JRA-55) dataset. The global topography data for the ocean model is derived from the Naval Research Laboratory Digital Bathymetry Data Base with a 2-minute resolution (NRL DBDB2). Additionally, seasonal climatic runoff data is incorporated(Fekete et al., 2002). The sea-air flux parameterization scheme implemented in this study follows the bulk formulas of Large and Pond (1981). The horizontal dissipation of momentum is handled using a Laplacian closure scheme with coefficients $\nu = 100 m^2 s^{-1}$, while the vertical dissipation includes

Laplace operators with coefficients $1 \times 10^{-3} m^2 s^{-1}$. The bottom layer employs a linear bottom drag coefficient of $2 \times 10^{-4} ms^{-1}$ (Fu et al., 2021). Due to computational constraints, the daily average simulation results for the period of 2019-2020 were calculated in this study.

Three experiments were conducted using the same initial condition to evaluate the performance of the CNN parameterization scheme. The first experiment, referred to as 9 km, involved a high-resolution simulation with a grid resolution of 9 km, serving as the reference simulation or "truth". The second experiment, referred to as 25 km, utilized a low-resolution simulation with a grid resolution of 25 km, without any parameterization. Finally, the third experiment, referred to as 25 km+CNN, employed a low-resolution simulation with a grid resolution of 25 km and incorporated the CNN mesoscale eddies parameterization. In the 25 km+CNN experiment, the velocities obtained from the low-resolution unparameterized experiment were used as inputs for the pre-trained CNN parameterization. Remarkably, the pre-trained CNN parameterization proved to be stable without the need for any further adjustments. It effectively generated global daily surface velocities with a grid resolution of 9 km, which were subsequently used for offline testing. By comparing the results obtained from these three experiments, we aim to assess the efficacy of the CNN parameterization scheme in enhancing the representation of mesoscale eddies under low-resolution conditions.

## 6.1 *Surface horizontal velocity and relative vorticity*

In this section of the study, we aimed to assess the sensitivity of different horizontal resolutions by comparing the 25 km+CNN parameterized simulations, the 9 km high-resolution simulation, and the 25 km low-resolution simulation without parameterization.

Figure 11 visualizes the surface horizontal velocity in the Northwest Pacific on January 1, 2019. The low-resolution experiment without parameterization displays lower surface velocity compared to the high-resolution experiment, specifically in the Kuroshio extension region. In this region, the maximum surface velocity is only 1.2 m/s. However, when the pre-trained CNN parameterization is directly applied to the low-resolution experiment, offline testing of surface velocity exhibits significant improvement due to mesoscale eddy-induced advection or transport. The maximum surface velocity increases to 1.8 m/s, which closely resembles the high-resolution experiment's value of 1.9 m/s. These results highlight the effectiveness of the CNN parameterization scheme in improving the representation of surface velocity in the low-resolution experiment and bringing it closer to the high-resolution simulation. This demonstrates the sensitivity of the model's performance to different horizontal resolutions and the potential of CNN parameterization in enhancing accuracy.

In addition to the constraints imposed on the learning process, the CNN parameterization scheme also greatly enhances the variance of the relative vorticity. In the low-resolution experiment (25 km), the eddies appear weakened, particularly in the central Northwest Pacific, to the point where they become almost indistinguishable. However, by incorporating the CNN parameterization in the 25 km+CNN experiment, the relative vorticity gradients are enhanced, allowing for a clearer identification of eddies that align with the high-resolution experiment (9 km). This improvement in the representation of eddies highlights the effectiveness of the CNN parameterization scheme in capturing and enhancing the variability of the relative vorticity field despite utilizing low-resolution inputs.

When compared to the 9 km experiment, the RMSE and relative error of the 25 km low-resolution test simulations from 2019 to 2020 amount to 4.7 cm and 21.3% respectively. Conversely, the 25 km+CNN parameterized simulations achieve more favorable results. The RMSE for these simulations is significantly reduced to 1.7 cm, and the relative error is also decreased to 8.4%. These findings suggest that the CNN parameterization scheme plays a crucial role in enhancing the accuracy and reducing errors in the simulations, bringing them closer to the performance of the high-resolution experiment.

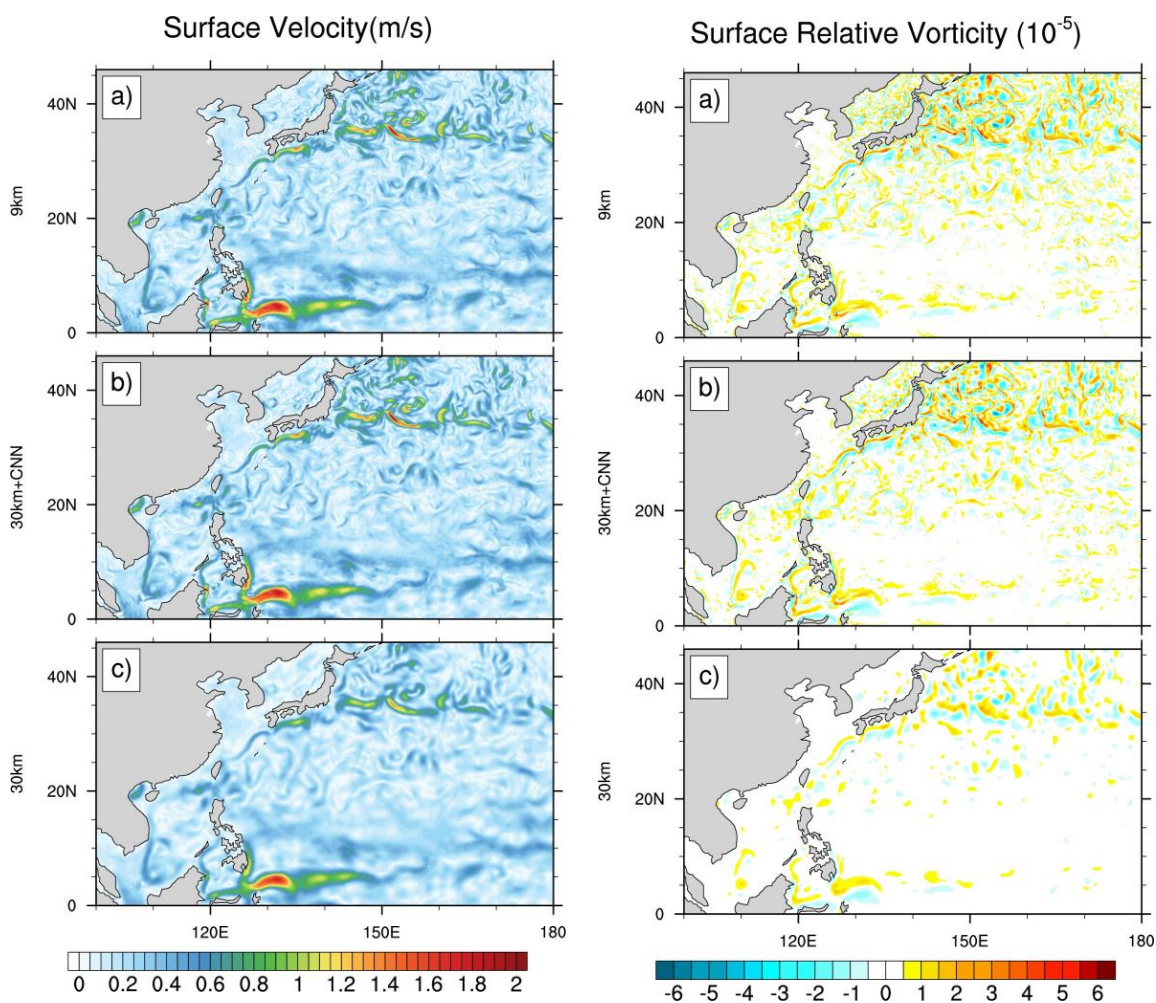

**Fig. 11.** The surface velocity field (left) and relative vorticity (right) for the three different experiments: (a) 9 km high-resolution simulation, (b) 25 km low-resolution

simulation with the CNN mesoscale eddies parameterization, and (c) 25 km low-resolution simulation without parameterization.

**6.2 *Eddies kinetic energy***

To investigate the variation patterns of EKE, we computed the average EKE in the Northwest Pacific. The surface EKE was calculated using the following method,

$$EKE = (u'^2 + v'^2)/2 \qquad (12)$$

where $u' = u - \bar{u}$ and $v' = v - \bar{v}$ are anomalous zonal and meridional components of surface geostrophic velocity, respectively. These components are calculated from sea level anomalies that are referenced to the mean sea level of 2019 and 2020. We calculate the daily EKE values by taking the square of the difference between the surface horizontal velocities their respective daily means, and then averaging over the entire grid points in the Northwest Pacific region.

The study proceeded to investigate the daily variations of EKE through different parameterization experiments. Figure 12 provides a visual representation of the distribution of the daily EKE in the Northwest Pacific. The plot clearly illustrates the presence of daily variations in EKE, with prominent peak values occurring during the summer season. These peak values indicate higher intensity of eddy activity during this time. Comparatively, the 25 km experiment demonstrates significantly weaker EKE values, measuring only about half of what is observed in the 9 km experiment. This suggests that the low-resolution simulation without parameterization fails to capture the same level of eddy activity as the high-resolution simulation.

On the other hand, both the 9 km experiment and the 25 km+CNN experiment exhibit similar maximum EKE intensities, reaching approximately 460 cm²/s². This indicates that the CNN parameterization is effective in enhancing the representation of EKE in the low-resolution experiment, bringing it closer to the high-resolution experiment. However, it is worth noting that the EKE values in the 25 km+CNN experiment are slightly lower than those in the 9 km experiment, suggesting that there may still be some small differences in the intensity of eddy activity between the two simulations.

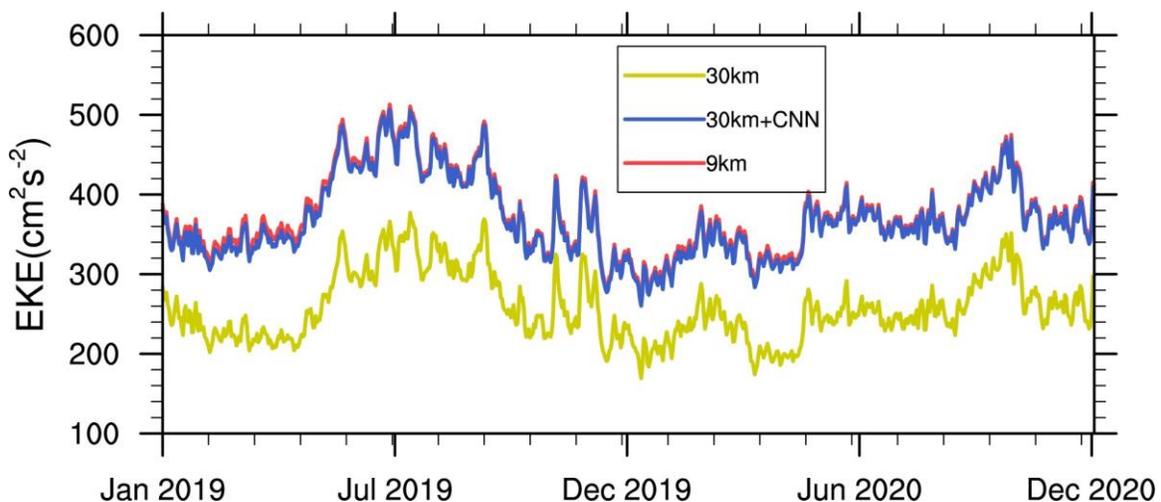

**Fig. 12.** The time series of the daily EKE in the Northwest Pacific is analyzed for the three different experiments: 9 km high-resolution simulation (red), 25 km low-resolution simulation with the CNN mesoscale eddies parameterization (blue) and 25 km low-resolution simulation without parameterization (yellow). The data is obtained from the test set and covers the period from 2019-2020.

Figure 13 showcases the standard deviations of EKE calculated over a two-year period (2019-2020) for the three experiments. In the 25 km experiment (Fig.13c), the surface velocity field displays weak intensity, and the eddies show insignificant time-varying characteristics. As a result, there is a considerable loss of variance in the Kuroshio and its extensions. This indicates that the low-resolution simulation without parameterization is unable to capture the full range of variability in the surface flow. In contrast, the 25 km+CNN experiment exhibits a substantial increase in EKE. This suggests that the CNN parameterization scheme has effectively improved the simulation of both the mean flow and the eddies. Consequently, the spatial and temporal variation characteristics of the surface flow observed in the 25 km+CNN experiment closely align with the results from the 9 km high-resolution experiment. This indicates that the CNN parameterization has successfully reduced the discrepancies in the simulation of surface flow between the low-resolution experiment and the high-resolution experiment.

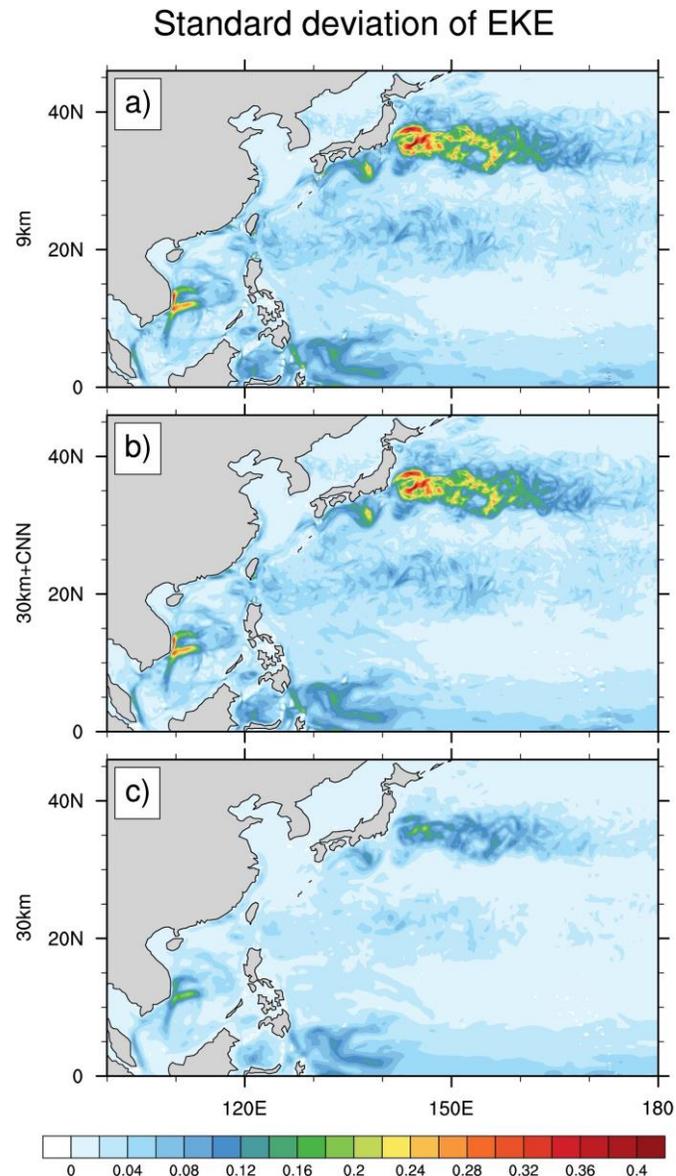

**Fig. 13.** Standard Deviation of the EKE in the Northwest Pacific for the three different experiments: (a) high-resolution simulation (9 km), (b) low-resolution simulation with the CNN mesoscale eddies parameterization and (c) low-resolution simulation (25 km) without parameterization. The data is obtained from the test set and covers the period from 2019-2020.

### 6.3 *Horizontal eddy kinetic energy spectrum density*

To calculate the horizontal kinetic energy spectral density, we first remove the temporal and spatial averages to obtain the horizontal spatio-temporal anomaly field. Then, we compute the two-dimensional (2D) discrete Fourier transform of the anomaly field and multiply it by a 2D Hanning window. The resulting spectrum is formed by multiplying the Fourier coefficients by their complex conjugate and averaging over all real numbers. We obtain the horizontal eddy kinetic energy spectral density by summing the spectral estimates of the two velocity components and dividing by two. We apply low and high wavelength cutoffs of 10 km and 500 km, respectively. This restricts the analysis to a specific range of wavelengths, allowing for a more focused investigation of the relevant energy scales.

Figure 14 depicts the comparison of the horizontal kinetic energy spectral densities for the three experiments. Overall, for wavelengths larger than 500 km, the slope of the kinetic energy spectral density remains relatively flat, indicating the presence of large-scale features that likely correspond to actual variations in the inertial range or dynamics. Both the low-resolution 25 km experiment and the high-resolution 9 km experiment exhibit these large-scale features.

Upon careful examination of the kinetic energy spectral densities (Fig.14), it becomes evident that the kinetic energy spectral densities of the 25 km experiment are significantly lower at all wavelengths compared to the 25 km+CNN experiment and the 9 km experiment. Furthermore, as the wave number increases, the kinetic energy spectral densities obtained from the 25 km experiment progressively deviate from those of the other two experiments. In conclusion, the kinetic energy spectral density of the 25 km experiment exhibits significantly reduced values across all wavelengths, indicating a less accurate representation of the energy distribution compared to the other two experiments.

The slope of the wavenumber spectrum of KE in the mesoscale band has been used to infer the dynamics of geostrophic oceanic flows (Xiao et al., 2023). The results of the kinetic energy spectral density slopes (Fig.14) demonstrate a close alignment between the 25 km+CNN and 9 km experiments within the wavelength range of 500 km to 100 km. Both experiments exhibit a $k^{-3}$ power law behavior in the kinetic energy spectral density slopes within the wavelength range of 200 km to 100 km, consistent with the predictions of two-dimensional quasi-geostrophic theory (Charney, 1971). In contrast, the 25 km experiment shows a significantly steeper slope in the kinetic energy spectral density. Despite using the unparameterized 25 km experiment results as input, the 25 km+CNN experiment achieves a similar level of kinetic energy spectral density as the high-resolution 9 km experiment across all spatial scales, implying a more effective backscatter or inverse energy cascade from the eddy to the mean flow (Richman et al., 2012).

Within the smaller wavelength range of 100 km to 20 km, both the 9 km experiment and the 25 km+CNN experiment exhibit a $k^{-5}$ power law behavior in the kinetic energy spectral density slopes. In the non-eddy-resolved 25 km experiment, the dissipation of eddy kinetic energy at small scales is spurious, hindering the compensation of energy from small to large scales (Jansen and Held, 2014). Both the 9 km experiment and the 25 km+CNN experiment have a notable impact in reducing energy dissipation at small scales. This reduction enables two-dimensional inverse energy compensation, wherein eddy kinetic energy is re-injected into the larger-scale flow field. This process significantly improves the kinetic energy spectral density across all wavenumbers, effectively enhancing the representation of energy distribution. The compensation enabled by the 25 km+CNN experiment play crucial roles in reducing energy dissipation, promoting inverse energy cascade, and enhancing the kinetic energy spectral density in the smaller wavelength range.

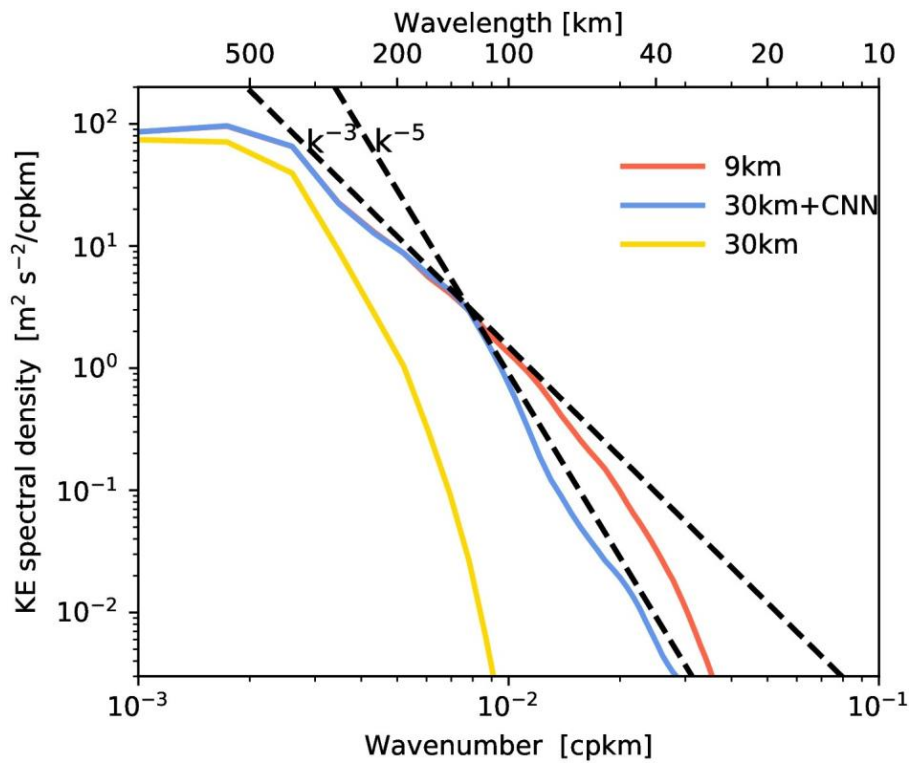

**Fig. 14.** The surface kinetic energy spectral density in the Northwest Pacific is analyzed for the three different experiments: 9 km high-resolution simulation (red), 25 km low-resolution simulation with the CNN mesoscale eddies parameterization (blue) and 25 km low-resolution simulation without parameterization (yellow). The data is obtained from the test set and covers the period from 2019-2020.

## 7. Conclusions

Subgrid eddies momentum and existing physical characteristic processes remain fundamental in parametric simulations, and deep learning techniques have the potential to assist in the exploration of new physical processes. This study presents a novel approach for capturing mesoscale eddies in ocean modeling by combining data-driven techniques and physical principles. The offline validation using ocean model simulations provides strong evidence of the effectiveness of the proposed CNN parameterization scheme. The findings contribute to enhancing the accuracy and reliability of ocean models in representing mesoscale eddies and their role in ocean dynamics and transport processes.

In this study, the CNN parameterization scheme used adheres to physics-based constraints, ensuring that it captures the spatio-temporal characteristics of subgrid eddy momentum accurately. The interpretability analysis conducted at both the global and example-based interpretation provides valuable insights into the relationship between the pre-trained CNN parameterization features and key known physical processes. The results of the interpretability analysis demonstrate a positive correlation between the features obtained from the pre-trained CNN parameterization and the underlying physical processes. This correlation indicates that the CNN parameterization effectively captures the important physical characteristics and behaviors of subgrid eddy momentum. Additionally, the interpretable CNN parameterization shows promising results in accurately solving the resolved mean velocity  at the local scale and effectively capturing the representation of unresolved subgrid turbulence processes at the global scale, demonstrates its potential for enhancing simulations and understanding of complex physical systems.

Unlike traditional fixed parameter estimation, the CNN parameterization scheme captures complex spatial and temporal variations of mesoscale eddies using characteristic parameters derived from a significant dataset. The results obtained demonstrate that the pre-trained CNN parameterization scheme only requires the low-resolution velocity field as input and significantly improves the simulation of mesoscale processes through kinetic energy backscatter. The integration of the CNN parameterization scheme within the MITgcm framework yields notable enhancements in the resolved mean flow and representation of mesoscale eddies across all wavenumbers. This improvement is observed even when using a low-resolution model.

**Declaration of competing interest**

The authors declare that they have no known competing financial interests or personal relationships that could have appeared to influence the work reported in this paper.

**Appendix A. CNN parameterization Training Hyperparameters**

To enhance the learning ability of the CNN parameterization scheme for nonlinear functions, we employ scaled exponential linear units (SELU) as the activation function in each convolution layer. The SELU activation function possesses self-normalization properties, allowing the network to converge towards zero mean and unit variance, even in the presence of noise and perturbations (Klambauer et al., 2017).

Furthermore, we tailor the network layer structure by adjusting various hyperparameters and evaluating the effectiveness of different configurations in minimizing the loss function of the validation dataset. This iterative process allows us to find the most suitable architecture for the CNN parameterization. The training of the CNN parameterization involves minimizing the soft constraint loss function of time back propagation. The optimizer used is Adam with a learning rate of 0.001 (Zeiler, 2012), which helps in efficiently updating the model parameters during the training process.

For more specific information regarding the hyperparameters utilized in training the CNN parameterization, please refer to Table A.1 for additional details. Please note that the exact values of the hyperparameters such as the number of convolution layers, filters per layer, kernel size, and neurons per dense layer may vary depending on the specific configuration that yielded the most effective results in minimizing the loss function of the validation dataset.

Overall, these strategies, including the employment of SELU activation function and the careful selection of network architecture and training parameters, aim to enhance the CNN's learning ability and ensure the effective modeling of the nonlinear functions involved in the parameterization scheme.

**Table A.1** Hyperparameters Used in the CNN Parameterization Scheme

| Hyperparameters | |
|---|---|
| **Inputs Size** | $2041 \times 4320 \times 2$ |
| **Output Size** | $2041 \times 4320 \times 1$ |
| **Number of training samples** | 1993-2019 (9495 days) |
| **Number of test samples** | 2019-2020 (731 days) |
| **Number of Convolution Layers** | 5 |
| **Number of Filters per Convolution Layer** | 32, 64, 128, 64, 32 |
| **Kernel Size of Convolution Layers** | $3 \times 3$ |
| **Activation Function** | SELU |
| **Loss Function** | Soft constraint Loss Function |
| **Optimizer** | Adam |
| **Learning Rate** | 0.001 |
| **Momentum** | 0.9 |
| **Batch Size** | 32 |
| **Epochs** | 100 |

**REFERENCES**

Abadi, M., Barham, P., Chen, J., Chen, Z., Davis, A., Dean, J., Devin, M., Ghemawat, S., Irving, G., Isard, M., 2016. {TensorFlow}: a system for {Large-Scale} machine learning, in: 12th USENIX Symposium on Operating Systems Design and Implementation (OSDI 16). pp. 265–283.

Abernathey, R.P., Marshall, J., 2013. Global surface eddy diffusivities derived from satellite altimetry. J. Geophys. Res. Ocean. 118, 901–916.

Alber, M., Lapuschkin, S., Seegerer, P., Hägele, M., Schütt, K.T., Montavon, G., Samek, W., Müller, K.-R., Dähne, S., Kindermans, P.-J., 2019. iNNvestigate neural networks! J. Mach. Learn. Res. 20, 1–8.

Aluie, H., Kurien, S., 2011. Joint downscale fluxes of energy and potential enstrophy in rotating stratified Boussinesq flows. Europhys. Lett. 96, 44006.

Bachman, S.D., 2019. The GM+E closure: A framework for coupling backscatter with the Gent and McWilliams parameterization. Ocean Model. 136, 85–106. https://doi.org/https://doi.org/10.1016/j.ocemod.2019.02.006

Barnes, E.A., Hurrell, J.W., Ebert-Uphoff, I., Anderson, C., Anderson, D., 2019. Viewing forced climate patterns through an AI lens. Geophys. Res. Lett. 46, 13389–13398.

Bauer, P., Thorpe, A., Brunet, G., 2015. The quiet revolution of numerical weather prediction. Nature 525, 47–55. https://doi.org/10.1038/nature14956

Berloff, P.S., 2005. Random-forcing model of the mesoscale oceanic eddies. J. Fluid Mech. 529, 71–95.

Bi, K., Xie, L., Zhang, H., Chen, X., Gu, X., Tian, Q., 2023. Accurate medium-range global weather forecasting with 3D neural networks. Nature. https://doi.org/10.1038/s41586-023-06185-3

Binder, A., Bach, S., Montavon, G., Müller, K.-R., Samek, W., 2016. Layer-wise relevance propagation for deep neural network architectures, in: Information Science and Applications (ICISA) 2016. Springer, pp. 913–922.

Bolton, T., Zanna, L., 2019. Applications of Deep Learning to Ocean Data Inference and Subgrid Parameterization. J. Adv. Model. Earth Syst. 11, 376–399. https://doi.org/10.1029/2018MS001472

Brenowitz, N.D., Bretherton, C.S., 2018. Prognostic Validation of a Neural Network Unified Physics Parameterization. Geophys. Res. Lett. 45, 6289–6298. https://doi.org/10.1029/2018GL078510

Charney, J.G., 1971. Geostrophic turbulence. J. Atmos. Sci. 28, 1087–1095.

Chelton, D.B., Schlax, M.G., Samelson, R.M., 2011. Global observations of nonlinear mesoscale eddies. Prog. Oceanogr. 91, 167–216. https://doi.org/10.1016/j.pocean.2011.01.002

Chen, Y., Zhang, D., 2020. Physics-constrained indirect supervised learning. Theor. Appl. Mech. Lett. 10, 155–160.

Cheng, L., Trenberth, K.E., Fasullo, J., Boyer, T., Abraham, J., Zhu, J., 2017. Improved estimates of ocean heat content from 1960 to 2015. Sci. Adv. 3, e1601545.

Chollet, F., 2015. keras.

Cybenko, G., 1989. Approximation by superpositions of a sigmoidal function. Math. Control. signals Syst. 2, 303–314.

Ding, M., Liu, H., Lin, P., Hu, A., Meng, Y., Li, Y., Liu, K., 2022. Overestimated Eddy Kinetic Energy in the Eddy-Rich Regions Simulated by Eddy-Resolving Global Ocean–Sea Ice Models. Geophys. Res. Lett. 49, e2022GL098370.

Ding, Z., Li, H., Zhou, D., Liu, Y., Hou, R., 2022. A robust infrared and visible image fusion framework via multi-receptive-field attention and color visual perception. Appl. Intell. 1–19.

Early, J.J., Samelson, R.M., Chelton, D.B., 2011. The evolution and propagation of quasigeostrophic ocean eddies. J. Phys. Oceanogr. 41, 1535–1555.

Ebert-Uphoff, I., Hilburn, K., 2020. Evaluation, tuning, and interpretation of neural networks for working with images in meteorological applications. Bull. Am. Meteorol. Soc. 101, E2149–E2170.

Eden, C., Greatbatch, R.J., 2008. Diapycnal mixing by meso-scale eddies. Ocean Model. 23, 113–120.

Eyring, V., Gleckler, P.J., Heinze, C., Stouffer, R.J., Taylor, K.E., Balaji, V., Guilyardi, E., Joussaume, S., Kindermann, S., Lawrence, B.N., Meehl, G.A., Righi, M., Williams, D.N., 2016. Towards improved and more routine Earth system model evaluation in CMIP. Earth Syst. Dyn. 7, 813–830. https://doi.org/10.5194/esd-7-813-2016

Farneti, R., Downes, S.M., Griffies, S.M., Marsland, S.J., Behrens, E., Bentsen, M., Bi, D., Biastoch, A., Boening, C., Bozec, A., 2015. An assessment of Antarctic Circumpolar Current and Southern Ocean meridional overturning circulation during 1958–2007 in a suite of interannual CORE-II simulations. Ocean Model. 93, 84–120.

Fekete, B.M., Vörösmarty, C.J., Grabs, W., 2002. High-resolution fields of global runoff combining observed river discharge and simulated water balances. Global Biogeochem. Cycles 16, 11–15.

Ferrari, R., Wunsch, C., 2009. Ocean circulation kinetic energy: Reservoirs, sources, and sinks. Annu. Rev. Fluid Mech. 41, 253–282.

Fu, H., Wu, X., Li, W., Zhang, L., Liu, K., Dan, B., 2021. Improving the accuracy of barotropic and internal tides embedded in a high-resolution global ocean circulation model of MITgcm. Ocean Model. 162, 101809.

Gagne, D.J., Christensen, H.M., Subramanian, A.C., Monahan, A.H., 2020. Machine learning for stochastic parameterization: Generative adversarial networks in the Lorenz'96 model. J. Adv. Model. Earth Syst. 12, e2019MS001896.

Gagne II, D.J., Haupt, S.E., Nychka, D.W., Thompson, G., 2019. Interpretable deep learning for spatial analysis of severe hailstorms. Mon. Weather Rev. 147, 2827–2845.

Gent, P.R., Mcwilliams, J.C., 1990. Isopycnal mixing in ocean circulation models. J. Phys. Oceanogr. 20, 150–155.

Gentine, P., Pritchard, M., Rasp, S., Reinaudi, G., Yacalis, G., 2018. Could Machine Learning Break the Convection Parameterization Deadlock? Geophys. Res. Lett. 45, 5742–5751. https://doi.org/10.1029/2018GL078202

Gerard, L., Piriou, J.-M., Brožková, R., Geleyn, J.-F., Banciu, D., 2009. Cloud and precipitation parameterization in a meso-gamma-scale operational weather prediction model. Mon. Weather Rev. 137, 3960–3977.

Ghahramani, Z., 2015. Probabilistic machine learning and artificial intelligence. Nature 521, 452–459. https://doi.org/10.1038/nature14541

Goodfellow, I., Bengio, Y., Courville, A., Bengio, Y., 2016. Deep learning. MIT press Cambridge.

Guillaumin, A.P., Zanna, L., 2021. Stochastic-deep learning parameterization of ocean momentum forcing. J. Adv. Model. Earth Syst. 13, e2021MS002534.

Hornik, K., Stinchcombe, M., White, H., 1989. Multilayer feedforward networks are universal approximators. Neural networks 2, 359–366.

Jansen, M.F., Held, I.M., 2014. Parameterizing subgrid-scale eddy effects using energetically consistent backscatter. Ocean Model. 80, 36–48.

Jean-Michel, L., Eric, G., Romain, B.-B., Gilles, G., Angélique, M., Marie, D., Clément, B., Mathieu, H., Olivier, L.G., Charly, R., 2021. The Copernicus global 1/12 oceanic and sea ice GLORYS12 reanalysis. Front. Earth Sci. 9, 698876.

Karniadakis, G.E., Kevrekidis, I.G., Lu, L., Perdikaris, P., Wang, S., Yang, L., 2021. Physics-informed machine learning. Nat. Rev. Phys. 3, 422–440.

Kim, J.H., Choi, J.H., Cheon, M., Lee, J.S., 2020. MAMNet: Multi-path adaptive modulation network for image super-resolution. Neurocomputing 402, 38–49. https://doi.org/10.1016/j.neucom.2020.03.069

Kindermans, P.J., Schütt, K.T., Alber, M., Müller, K.R., Erhan, D., Kim, B., Dähne, S., 2018. Learning how to explain neural networks: Patternnet and Patternattribution. 6th Int. Conf. Learn. Represent. ICLR 2018 - Conf. Track Proc.

Klambauer, G., Unterthiner, T., Mayr, A., Hochreiter, S., 2017. Self-normalizing neural networks. Adv. Neural Inf. Process. Syst. 30.

Krishnan, M., 2020. Against interpretability: a critical examination of the interpretability problem in machine learning. Philos. Technol. 33, 487–502.

Lapuschkin, S., Wäldchen, S., Binder, A., Montavon, G., Samek, W., Müller, K.-R., 2019. Unmasking Clever Hans predictors and assessing what machines really learn. Nat. Commun. 10, 1–8.

Lee, Y.-J., Hall, D., Stewart, J., Govett, M., 2018. Machine learning for targeted assimilation of satellite data, in: Joint European Conference on Machine Learning and Knowledge Discovery in Databases. Springer, pp. 53–68.

Li, G., Yu, Y., 2016. Visual saliency detection based on multiscale deep CNN features. IEEE Trans. image Process. 25, 5012–5024.

Li, H., Storch, J.-S. von, 2013. On the fluctuating buoyancy fluxes simulated in a OGCM. J. Phys. Oceanogr. 43, 1270–1287.

Li, X., Liu, B., Zheng, G., Ren, Y., Zhang, S., Liu, Yingjie, Gao, L., Liu, Yuhai, Zhang, B., Wang, F., 2020. Deep learning-based information mining from ocean remote sensing imagery. Natl. Sci. Rev. https://doi.org/10.1093/nsr/nwaa047

Mak, J., Maddison, J.R., Marshall, D.P., Munday, D.R., 2018. Implementation of a geometrically informed and energetically constrained mesoscale eddy parameterization in an ocean circulation model. J. Phys. Oceanogr. 48, 2363–2382.

Mana, P.P., Zanna, L., 2014. Toward a stochastic parameterization of ocean mesoscale eddies. Ocean Model. 79, 1–20.

Marshall, D.P., Maddison, J.R., Berloff, P.S., 2012. A framework for parameterizing eddy potential vorticity fluxes. J. Phys. Oceanogr. 42, 539–557.

Marshall, J., Adcroft, A., Hill, C., Perelman, L., Heisey, C., 1997a. A finite-volume, incompressible Navier Stokes model for studies of the ocean on parallel computers.

J. Geophys. Res. Ocean. 102, 5753–5766.
https://doi.org/https://doi.org/10.1029/96JC02775

Marshall, J., Hill, C., Perelman, L., Adcroft, A., 1997b. Hydrostatic, quasi-hydrostatic, and nonhydrostatic ocean modeling. J. Geophys. Res. Ocean. 102, 5733–5752. https://doi.org/https://doi.org/10.1029/96JC02776

Maximenko, N.A., Bang, B., Sasaki, H., 2005. Observational evidence of alternating zonal jets in the world ocean. Geophys. Res. Lett. 32. https://doi.org/https://doi.org/10.1029/2005GL022728

McGovern, A., Lagerquist, R., Gagne, D.J., Jergensen, G.E., Elmore, K.L., Homeyer, C.R., Smith, T., 2019. Making the black box more transparent: Understanding the physical implications of machine learning. Bull. Am. Meteorol. Soc. 100, 2175–2199.

Molnar, C., 2020. Interpretable machine learning. Lulu. com.

Montavon, G., Lapuschkin, S., Binder, A., Samek, W., Müller, K.R., 2017. Explaining nonlinear classification decisions with deep Taylor decomposition. Pattern Recognit. 65, 211–222. https://doi.org/10.1016/J.PATCOG.2016.11.008

Montavon, G., Samek, W., Müller, K.R., 2018. Methods for interpreting and understanding deep neural networks. Digit. Signal Process. 73, 1–15. https://doi.org/10.1016/J.DSP.2017.10.011

Pathak, J., Subramanian, S., Harrington, P., Raja, S., Chattopadhyay, A., Mardani, M., Kurth, T., Hall, D., Li, Z., Azizzadenesheli, K., 2022. FourCastNet: A Global Data-driven High-resolution Weather Model using Adaptive Fourier Neural Operators. arXiv Prepr. arXiv2202.11214.

Pope, Sb., 1975. A more general effective-viscosity hypothesis. J. Fluid Mech. 72, 331–340.

Raissi, M., Yazdani, A., Karniadakis, G.E., 2020. Hidden fluid mechanics: Learning velocity and pressure fields from flow visualizations. Science (80-. ). 367, 1026–1030.

Richman, J.G., Arbic, B.K., Shriver, J.F., Metzger, E.J., Wallcraft, A.J., 2012. Inferring dynamics from the wavenumber spectra of an eddying global ocean model with embedded tides. J. Geophys. Res. Ocean. 117.

Rudin, C., 2019. Stop explaining black box machine learning models for high stakes decisions and use interpretable models instead. Nat. Mach. Intell. 1, 206–215.

Samek, W., Montavon, G., Vedaldi, A., Hansen, L.K., Müller, K.-R., 2019. Explainable AI: interpreting, explaining and visualizing deep learning. Springer Nature.

Schneider, T., Kaul, C.M., Pressel, K.G., 2019. Possible climate transitions from breakup of stratocumulus decks under greenhouse warming. Nat. Geosci. 12, 163–167.

Simonyan, K., Vedaldi, A., Zisserman, A., 2013. Deep inside convolutional networks: Visualising image classification models and saliency maps. arXiv Prepr. arXiv1312.6034.

Smagorinsky, J., 1963. General circulation experiments with the primitive equations: I. The basic experiment. Mon. Weather Rev. 91, 99–164.

Smilkov, D., Thorat, N., Kim, B., Viégas, F., Wattenberg, M., 2017. Smoothgrad: removing noise by adding noise. arXiv Prepr. arXiv1706.03825.

Toms, B.A., Barnes, E.A., Ebert-Uphoff, I., 2020. Physically Interpretable Neural Networks for the Geosciences: Applications to Earth System Variability. J. Adv.

Model. Earth Syst. 12, e2019MS002002.
https://doi.org/https://doi.org/10.1029/2019MS002002

Wang, G., Wang, X., Wu, X., Liu, K., Qi, Y., Sun, C., Fu, H., 2021. A Hybrid Multivariate Deep Learning Networks for Multistep ahead Sea Level Anomaly Forecasting. J. Atmos. Ocean. Technol. https://doi.org/10.1175/JTECH-D-21-0043.1

Wang, P., Jiang, J., Lin, P., Ding, M., Wei, J., Zhang, F., Zhao, L., Li, Y., Yu, Z., Zheng, W., 2021. The GPU version of LASG/IAP Climate System Ocean Model version 3 (LICOM3) under the heterogeneous-compute interface for portability (HIP) framework and its large-scale application. Geosci. Model Dev. 14, 2781–2799.

Waterman, S., Hoskins, B.J., 2013. Eddy shape, orientation, propagation, and mean flow feedback in western boundary current jets. J. Phys. Oceanogr. 43, 1666–1690.

Xiao, B., Qiao, F., Shu, Q., Yin, X., Wang, G., Wang, S., 2023. Development and validation of a global 1⁄32° surface-wave–tide–circulation coupled ocean model: FIO-COM32. Geosci. Model Dev. 16, 1755–1777. https://doi.org/10.5194/gmd-16-1755-2023

Xu, K., Darve, E., 2022. Physics constrained learning for data-driven inverse modeling from sparse observations. J. Comput. Phys. 453, 110938.
https://doi.org/10.1016/j.jcp.2021.110938

Zeiler, M.D., 2012. ADADELTA: An Adaptive Learning Rate Method.

Zhang, Z., Liu, Yuelin, Qiu, B., Luo, Y., Cai, W., Yuan, Q., Liu, Yinxing, Zhang, H., Liu, H., Miao, M., Zhang, J., Zhao, W., Tian, J., 2023. Submesoscale inverse energy cascade enhances Southern Ocean eddy heat transport. Nat. Commun. 14, 1335.
https://doi.org/10.1038/s41467-023-36991-2

Zheng, G., Li, X., Zhang, R.H., Liu, B., 2020. Purely satellite data-driven deep learning forecast of complicated tropical instability waves. Sci. Adv. 6, eaba1482. https://doi.org/10.1126/sciadv.aba1482

Zhu, Y., Zhang, R.-H., Moum, J.N., Wang, F., Li, X., Li, D., 2022. Physics-informed deep learning parameterization of ocean vertical mixing improves climate simulations. Natl. Sci. Rev.

*Acknowledgments.* We would like to express our gratitude to Prof. Yiquan Qi for providing valuable advice and guidance throughout the course of this work. Additionally, we would like to acknowledge Dr. Caixia Shao for her assistance with the fit computation. This research was financially supported by the National Key Research and Development Program of China under grant number 2021YFC3101602 and the National Natural Science Foundation of China under grant numbers 42176017 and 41976019. We would also like to extend our appreciation to the Copernicus Marine Service for providing access to the Copernicus Marine Environment Monitoring Service global ocean 1/12° physical reanalysis GLORYS12V1 dataset (available at https://marine.copernicus.eu). The availability of this dataset was instrumental in carrying out our analysis and advancing our understanding in this field.